\documentclass[preprint,aps]{revtex4}

\newcommand{\cC}{\ensuremath{\mathcal{C}}}
\newcommand{\cP}{\ensuremath{\mathcal{P}}}
\newcommand{\cT}{\ensuremath{\mathcal{T}}}
\newcommand{\cQ}{\ensuremath{\mathcal{Q}}}

\begin{document}

\title{Exactly solvable $\cP\cT$-symmetric Hamiltonian having no Hermitian
counterpart}

\author{Carl~M.~Bender${}^1$ and Philip~D.~Mannheim${}^2$}

\affiliation{${}^1$Physics Department\\ Washington University\\ St.~Louis, MO
63130, USA\\ {\tt electronic address: cmb@wustl.edu}\\ \\
${}^2$Department of Physics\\ University of Connecticut\\ Storrs, CT
06269, USA
\\{\tt electronic address: philip.mannheim@uconn.edu}}

\date{April 24, 2008}

\begin{abstract}
In a recent paper Bender and Mannheim showed that the unequal-frequency
fourth-order derivative Pais-Uhlenbeck oscillator model has a realization in
which the energy eigenvalues are real and bounded below, the Hilbert-space
inner
product is positive definite, and time evolution is unitary. Central to that
analysis was the recognition that the Hamiltonian $H_{\rm PU}$ of the
model is
$\cP\cT$ symmetric. This Hamiltonian was mapped to a conventional
Dirac-Hermitian Hamiltonian via a similarity transformation whose form was
found
exactly. The present paper explores the equal-frequency limit of the same
model. It is shown that in this limit the similarity transform that was used
for the unequal-frequency case becomes singular and that $H_{\rm PU}$
becomes a
Jordan-block operator, which is nondiagonalizable and has fewer energy
eigenstates than eigenvalues. Such a Hamiltonian has no Hermitian
counterpart.
Thus, the equal-frequency $\cP\cT$ theory emerges as a distinct
realization of
quantum mechanics. The quantum mechanics associated with this Jordan-block
Hamiltonian can be treated exactly. It is shown that the Hilbert space is
complete with a set of nonstationary solutions to the Schr\"odinger equation
replacing the missing stationary ones. These nonstationary states are
needed to
establish that the Jordan-block Hamiltonian of the equal-frequency
Pais-Uhlenbeck model generates unitary time evolution.
\end{abstract}

\maketitle

\section{Introduction}
\label{s0}

A decade ago it was discovered that the non-Dirac-Hermitian Hamiltonian
$H=p^2+
ix^3$ has an entirely real quantum-mechanical energy spectrum \cite{R1}. The
reason for this unexpected spectral reality is that this Hamiltonian has an
underlying $\cP \cT$ symmetry and that this symmetry is {\it unbroken};
that is,
the energy eigenstates are also eigenstates of the $\cP\cT$ operator. In
general, whenever a Hamiltonian has an unbroken $\cP\cT$ symmetry, its energy
spectrum is real \cite{R2,R3}. Unbroken $\cP\cT$ invariance serves as an
alternative to Dirac Hermiticity in quantum theory. Moreover, for any
non-Hermitian Hamiltonian that has a complete basis of real-energy
eigenstates,
there necessarily exists a similarity transformation that brings it to
diagonal
Hermitian form. Because the transformation is a similarity rather than a
unitary
one, such unbroken-$\cP\cT$-symmetric Hamiltonians are really Dirac-Hermitian
Hamiltonians written in a disguised form in terms of a skew basis.
However, even
if a non-Hermitian Hamiltonian has a real eigenspectrum, it is not
automatically
diagonalizable because it might be of Jordan-block form. In such an event no
similarity transformation exists and the $\cP\cT$ sector then emerges as a
distinct and self-contained realization of quantum mechanics.

In this paper we study an exactly solvable model, namely, the equal-frequency
version of the Pais-Uhlenbeck oscillator \cite{R4} in which a distinct
$\cP\cT$
realization arises and for which the $\cP\cT$-symmetric Hamiltonian has no
Dirac-Hermitian counterpart. We explore the quantum mechanics associated with
such a self-contained $\cP\cT$ quantum realization and give special
attention to
the lack of completeness of the energy eigenstates that is characteristic of
Jordan-block Hamiltonians. Previously, we were motivated to study the
unequal-frequency version of the Pais-Uhlenbeck model because of its ghost
problem \cite{R5}. Here, we study the Jordan-block structure of the
equal-frequency model by extending the techniques that were developed in
Ref.~\cite{R5} to resolve the ghost problem in the unequal-frequency
version of
the model. We show that even though the Pais-Uhlenbeck-model Hamiltonian
develops Jordan-block structure in the equal-frequency limit, the
unitarity of
the theory is not lost.

This paper is organized as follows: Section~\ref{s1} provides a brief
summary of
$\cP\cT$ quantum mechanics and Sec.~\ref{s2} reviews the results of
Ref.~\cite{R5} for the unequal-frequency Pais-Uhlenbeck model. In
Sec.~\ref{s3}
we construct the Fock space associated with the unequal-frequency
Pais-Uhlenbeck
model. Then, in Sec.~\ref{s4} we use the results of Sec.~\ref{s3} to
construct
the equal-frequency Fock space, and we show that a Jordan-block form for the
Hamiltonian with its incomplete set of energy eigenstates arises in the
equal-frequency limit. In Sec.~\ref{s5} we construct the eigenfunctions of
the
unequal-frequency theory, and in Sec.~\ref{s6} we take their limit to
construct
the eigenfunctions of the equal-frequency theory. In doing so, we discover
what happened to the missing energy eigenstates. In Sec.~\ref{s7} we present
some conclusions and comments. Finally, in the Appendix we discuss the
implementation of the Lehmann representation in theories having non-Hermitian
Hamiltonians.

\section{Brief Summary of $\cP\cT$ Quantum Mechanics}
\label{s1}

There has been much research during the past few years on $\cP\cT$ quantum
mechanics; early references include Refs.~\cite{R1,R2,R6} and some recent
reviews may be found in Refs.~\cite{R3,R7}. A $\cP\cT$ quantum theory is
one whose dynamics is governed by a Hamiltonian $H$ that commutes with the
$\cP\cT$ operator. (Here, $\cP$ is the parity operator, which performs
spatial
reflection, and $\cT$ is the time-reversal operator.) The appeal of $\cP\cT$
quantum mechanics is that even if a Hamiltonian is not Dirac Hermitian it
will
still have real energy eigenvalues whenever it is $\cP\cT$ symmetric and
all of
its eigenstates are also eigenstates of the $\cP\cT$ operator.

For diagonalizable $\cP\cT$-symmetric Hamiltonians it is convenient to
construct
an operator $\cC$, which obeys the three simultaneous algebraic operator
equations
\begin{equation}
\cC^2={\bf 1},\quad[\cC,\cP\cT]=0,\quad [\cC,H]=0.
\label{K1}
\end{equation}
The first two of these equations are kinematical, while the third is
dynamical
because it involves the Hamiltonian $H$. In terms of $\cC$ there is a formal
construction of an operator $e^\cQ$:
\begin{equation}
e^{\cQ}=\cC\cP,
\label{K2}
\end{equation}
where the operator $\cQ$ is Hermitian in the conventional Dirac sense
\cite{R3}. Using the operator $\cQ$ it is possible (at least in principle) to
map the Hamiltonian $H$ to a Dirac-Hermitian Hamiltonian $\tilde H$ by
means of
a similarity transformation of the form \cite{R8,R9}:
\begin{equation}
{\tilde H}=e^{-\cQ/2}He^{\cQ/2}.
\label{K3}
\end{equation}

While the similarity transformation (\ref{K3}) is isospectral, it is not
unitary
because $\cQ$ is Hermitian rather than anti-Hermitian. Hence, while
$\tilde H$
and $H$ have the same energy eigenvalues $E_n$, their energy eigenkets are
not
unitarily equivalent. Rather, the eigenkets $|{\tilde n}\rangle$ and
$|n\rangle$
of $\tilde H$ and $H$ satisfy
\begin{equation}
{\tilde H}|{\tilde n}\rangle=E_n|{\tilde n}\rangle
\label{K4}
\end{equation}
and
\begin{equation}
H|n\rangle=E_n|n\rangle,
\label{K5}
\end{equation}
and are related by the mapping
\begin{equation}
|n\rangle=e^{\cQ/2}|{\tilde n}\rangle.
\label{K6}
\end{equation}

The energy eigenbra states corresponding to these eigenkets cannot be
obtained from the kets by simple Dirac conjugation. To construct the eigenbra
states of $H$ we take the Dirac-Hermitian conjugate of (\ref{K4}):
\begin{equation}
\langle{\tilde n}|{\tilde H}=E_n\langle{\tilde n}|.
\label{K7}
\end{equation}
We then define
\begin{equation}
\langle n|\equiv\langle{\tilde n}|e^{\cQ/2}
\label{K8}
\end{equation}
and note that $\langle n|$ is not an eigenbra state of $H$. Rather, the
eigenbra
state of $H$ is given by
\begin{equation}
\langle n|e^{-\cQ}H=\langle n|e^{-\cQ}E_n.
\label{K9}
\end{equation}

The eigenbra and eigenket states of the Dirac-Hermitian Hamiltonian
$\tilde H$
obey the usual statements of orthogonality, completeness, and Hamiltonian
operator reconstruction:
\begin{equation}
\langle{\tilde n}|{\tilde m}\rangle=\delta_{m,n},
\label{K10}
\end{equation}
\begin{equation}
\sum_n|{\tilde n}\rangle\langle{\tilde n}|={\bf 1},
\label{K11}
\end{equation}
\begin{equation}
{\tilde H}=\sum_n|{\tilde n}\rangle E_n\langle{\tilde n}|.
\label{K12}
\end{equation}
The similarity transformations in (\ref{K6}) and (\ref{K8}) imply that for
the
non-Hermitian Hamiltonian $H$ the corresponding statements are
\begin{equation}
\langle n|e^{-\cQ}|m\rangle=\delta_{m,n},
\label{K13}
\end{equation}
\begin{equation}
\sum_n|n\rangle\langle n|e^{-\cQ}={\bf 1},
\label{K14}
\end{equation}
\begin{equation}
H=\sum_n|n\rangle E_n\langle n|e^{-\cQ}.
\label{K15}
\end{equation}

The norm in (\ref{K13}) is relevant for the $\cP\cT$-symmetric Hamiltonian
$H$,
with $\langle n|e^{-\cQ}$ rather than $\langle n|$ being the appropriate
energy
eigenbra. In any $\cP\cT$ theory for which the operator $e^{-\cQ}$ exists,
there
will be a positive norm of the form given in (\ref{K13}) and no states of
negative norm (ghost states). Furthermore, because $[H,\cC\cP\cT]=0$, the
$\cP
\cT$-symmetric Hamiltonian $H$ will generate unitary time evolution even
though
it is not Hermitian.

To underscore the need for a non-Dirac norm for non-Hermitian
Hamiltonians, we
recall the connection between the Schr\"odinger and Heisenberg
representations.
Specifically, in the Schr\"odinger representation one introduces
time-dependent
states and time-independent operators. Ordinarily one does this for a
Hermitian
Hamiltonian, and it is instructive to see how things change in the
non-Hermitian
case. We thus consider Schr\"odinger equations for ket and bra states:
\begin{equation}
i\frac{d}{dt}|\alpha_{\rm S}(t)\rangle=H|\alpha_{\rm S}(t)\rangle,\qquad
-i\frac{d}{dt}\langle\alpha_{\rm S}(t)|=\langle\alpha_{\rm S}(t)|H^\dag.
\label{K16}
\end{equation}
For time-independent $H$ and $H^\dag$ the solutions to these equations are
\begin{equation}
|\alpha_{\rm S}(t)\rangle=e^{-iHt}|\alpha_{\rm S}(0)\rangle,\qquad
\langle\alpha_{\rm S}(t)|=\langle\alpha_{\rm S}(0)|e^{iH^\dag t}.
\label{K17}
\end{equation}

We introduce a time-independent (Schr\"odinger) operator $A_{\rm S}$ with
matrix
element
\begin{equation}
\langle \alpha_{\rm S}(t)|A_{\rm S}|\alpha_{\rm S}(t)\rangle = \langle
\alpha_{\rm S}(0)|e^{iH^{\dagger}t}A_{\rm S}e^{-iHt}|\alpha_{\rm
S}(0)\rangle,
\label{K18}
\end{equation}
and define the time-dependent (Heisenberg) operator
\begin{equation}
A_{\rm H}(t)=e^{iH^\dag t}A_{\rm S}e^{-iHt}.
\label{K19}
\end{equation}
The operator $A_{\rm H}(t)$ obeys
\begin{equation}
i\frac{d}{dt}A_{\rm H}(t)=A_{\rm H}(t)H-H^{\dagger}A_{\rm H}(t).
\label{K20}
\end{equation}
Since $H$ and $H^\dag$ are different when $H$ is not Dirac Hermitian, the
time
derivative of $A_{\rm H}(t)$ is not given by the commutator of $A_{\rm H}(t)$
with $H$. However, regardless of whether or not the Hamiltonian is Dirac
Hermitian, the
Hamiltonian {\it is} the generator of time translations. Thus, in the
Heisenberg
representation the time derivative of an operator is always given by
\begin{equation}
i\frac{d}{dt}A_{\rm H}(t)=A_{\rm H}(t)H-HA_{\rm H}(t).
\label{K21}
\end{equation}

To construct the Schr\"odinger states we start from (\ref{K21}) and work
backward. Thus, we replace (\ref{K16}) and (\ref{K17}) by
\begin{equation}
i\frac{d}{dt}|\alpha_{\rm S}(t)\rangle=H|\alpha_{\rm S}(t)\rangle,\qquad
-i\frac
{d}{dt}\langle\hat{\alpha}_{\rm S}(t)|=\langle\hat{\alpha}_{\rm S}(t)|H,
\label{K22}
\end{equation}
so
\begin{equation}
|\alpha_{\rm S}(t)\rangle=e^{-iHt}|\alpha_{\rm S}(0)\rangle,\qquad
\langle\hat{\alpha}_{\rm S}(t)|=\langle\hat{\alpha}_{\rm S}(0)|e^{iHt},
\label{K23}
\end{equation}
where the bra state $\langle \hat{\alpha}_{\rm S}(t)|$ is not the Dirac
conjugate of the ket state $|\alpha_{\rm S}(t)\rangle$. In (\ref{K22}) we see
that $H$ acts to the right on $|\alpha_{\rm S}(t)\rangle$ and to the left on
$\langle\hat{\alpha}_{\rm S}(t)|$. The appropriate inner product is given by
$\langle\hat{\beta}_{\rm S}(t)|\alpha_{\rm S}(t)\rangle$. Because of
(\ref{K23}) we have
\begin{equation}
\langle\hat{\beta}_{\rm S}(t)|\alpha_{\rm S}(t)\rangle=
\langle\hat{\beta}_{\rm S}(0)|e^{iHt}e^{-iHt}|\alpha_{\rm S}(0)\rangle=
\langle\hat{\beta}_{\rm S}(0)|\alpha_{\rm S}(0)\rangle,
\label{K24}
\end{equation}
which is the statement of unitary time development. By contrast, the Dirac
inner
product
\begin{equation}
\langle\beta_{\rm S}(t)|\alpha_{\rm S}(t)\rangle=\langle\beta_{\rm S}(0)|
e^{iH^\dag t}e^{-iHt}|\alpha_{\rm S}(0)\rangle\neq\langle\beta_{\rm S}(0)
|\alpha_{\rm S}(0)\rangle
\label{K25}
\end{equation}
is not time independent. Thus, for non-Hermitian Hamiltonians unitarity is
achieved by using an inner product that is different from the usual Dirac
inner
product. For $\cP\cT$-symmetric Hamiltonians the states $\langle\beta_{\rm
S}(t)
|$ and $\langle\hat{\beta}_{\rm S}(t)|$ are related by
$\langle\hat{\beta}_{\rm
S}(t)|=\langle\beta_{\rm S}(t)|e^{-\cQ}$.

Several models have been discussed in the literature for which one can
calculate
the operator $\cQ$ in closed form. Amongst them is the Lee model \cite{R10},
where by calculating the correct inner product one can show that the model is
explicitly ghost free \cite{R11}. Examples in which one can find exact
expressions for the equivalent Dirac-Hermitian Hamiltonian $\tilde H$ in
(\ref{K3}) associated with a given $\cP\cT$-symmetric Hamiltonian may be
found
in Refs.~\cite{R12,R13,R14,R15,R16} and in our work in Ref.~\cite{R5}.

We emphasize that if we are given a Dirac-Hermitian Hamiltonian $\tilde H$
and
we convert it to non-Hermitian form $H$ by means of the similarity
transformation $H=e^{\cQ/2}{\tilde H}e^{-\cQ/2}$, we know that we are dealing
with a Hermitian Hamiltonian in disguise. However, if we start with a
non-Hermitian Hamiltonian $H$, we do not immediately know if $H$ is a
disguised
Dirac-Hermitian Hamiltonian, and the advantage of $\cP\cT$ symmetry is
that it
provides a diagnostic for determining whether this might in fact be the
case. It
may happen that the operator $\cQ$ simply does not exist, and when this is
the
case, the $\cP\cT$-symmetric Hamiltonian will have no Hermitian counterpart.
This situation arises when $H$ has some complex eigenvalues (its $\cP\cT$
symmetry is broken), and thus there is obviously no Hermitian counterpart. A
transition from an unbroken to a broken $\cP\cT$ symmetry has actually been
observed in recent laboratory optics experiments \cite{R17}.

However, in this paper we encounter a more serious and fundamental
obstacle in
trying to construct a Hermitian Hamiltonian $\tilde H$ associated with a
non-Hermitian $\cP\cT$-symmetric Hamiltonian $H$. Specifically, even if the
eigenvalues of $H$ are all real, $H$ may be a Jordan-block matrix that has
fewer
eigenfunctions than eigenvalues, and thus it is not diagonalizable
\cite{R18}.
This is the case with the equal-frequency Pais-Uhlenbeck oscillator model.
For
this case the operator $\cQ$ of the unequal-frequency oscillator model
becomes
singular in the equal-frequency limit.

\section{Review of the Unequal-Frequency Pais-Uhlenbeck Oscillator Model}
\label{s2}

The Pais-Uhlenbeck oscillator was introduced in 1950 as a simple model to
explore the structure of a quantum system whose Lagrangian depends on
acceleration as well as on position and velocity \cite{R4}. The
Pais-Uhlenbeck
action is
\begin{equation}
I_{\rm PU}=\frac{\gamma}{2}\int dt\left[{\ddot z}^2-\left(\omega_1^2
+\omega_2^2\right){\dot z}^2+\omega_1^2\omega_2^2z^2\right],
\label{K26}
\end{equation}
where $\gamma$, $\omega_1$, and $\omega_2$ are all positive constants.
Because
the action depends on the acceleration, the differential equation of motion
\begin{equation}
\frac{d^4z}{dt^4}+(\omega_1^2+\omega_2^2)\frac{d^2z}{dt^2}+\omega_1^2\omega_2^2
z=0
\label{K27}
\end{equation}
is fourth order.

The Pais-Uhlenbeck oscillator model is interesting because a fourth-order
wave
equation leads to a Green's function $G(E)$ whose denominator in energy
space is
quartic in the energy. To find the Green's function, we replace the right
side
of (\ref{K27}) by a delta-function source term and take the Fourier
transform.
The result is
\begin{equation}
G(E)=\frac{1}{(E^2-\omega_1^2)(E^2-\omega_2^2)}.
\label{K28}
\end{equation}
The advantage of such a Green's function is that it leads to Feynman
integrals
that are more convergent than the corresponding integrals constructed from
propagators having quadratic denominators.

One may worry that there is a price to pay for such good convergence
because the
form of this Green's function seems to imply the existence of a ghost
state when
$\omega_1\neq\omega_2$. The argument goes as follows: In partial fraction
form
the Green's function
\begin{equation}
G(E)=\frac{1}{\omega_1^2-\omega_2^2}\left(\frac{1}{E^2-\omega_1^2}-\frac{1}{E^2-
\omega_2^2}\right)
\label{K29}
\end{equation}
describes the propagation of two kinds of states, one of energy $\omega_1$
and
the other of energy $\omega_2$. Assuming without loss of generality that
$\omega_1>\omega_2$, it appears that $\omega_2$ is associated with a state of
{\it negative} probability because its residue contribution to the
propagator is
negative. This appears to violate the positivity condition on the weight
function of the Lehmann representation. (Recall that when the two-point
Green's
function is expressed in Lehmann-representation form the requirement that all
quantum states have positive Dirac norm implies that the residues of all
intermediate propagating states must be strictly positive. See, for example,
Ref.~\cite{R19}.) In the past, theories having fourth-order wave equations
have
been abandoned because they were thought to violate the
Lehmann-representation
positivity condition.

However, the Pais-Uhlenbeck oscillator was recently revisited \cite{R5}
and it
was shown that there is a realization of the unequal-frequency model in which
the Hilbert space actually contains no ghost states. Specifically, in
Ref.~\cite{R5} it was shown that the above Green's-function argument has a
subtle flaw, namely the presumption that the Hamiltonian for the model
\cite{R20,R21,R22}
\begin{equation}
H_{\rm PU}=\frac{p_x^2}{2\gamma}+p_zx+\frac{\gamma}{2}\left(\omega_1^2
+\omega_2^2 \right)x^2-\frac{\gamma}{2}\omega_1^2\omega_2^2z^2
\label{K30}
\end{equation}
is Dirac Hermitian and that its associated norm is the standard Dirac norm.
[Because the action of (\ref{K26}) is constrained \cite{R20,R21,R22}, to
construct the Hamiltonian it was necessary to replace $\dot{z}$ by a new and
independent variable $x$ in the action (\ref{K26}).] It was shown in
Ref.~\cite{R5} that one should interpret the Hamiltonian $H_{\rm PU}$ as a
member of the class of non-Dirac-Hermitian Hamiltonians that are symmetric
under
combined space reflection $\cP$ and time reversal $\cT$. Thus, as
explained in
Sec.~\ref{s1}, it is necessary to replace the Dirac inner product by the
inner
product that we argue is appropriate for the Pais-Uhlenbeck Hamiltonian.

When we use the norm in (\ref{K13}) for the unequal-frequency Pais-Uhlenbeck
model, this model becomes ghost free and unitary. Moreover, because the
norm in
(\ref{K13}) is not the conventional Dirac norm, the relative negative sign
that
appears in the Lehmann representation in (\ref{K29}) cannot be interpreted as
the residue associated with a negative-norm (ghost) state. Rather, we show
below
that this negative sign is associated with an eigenvalue of the $\cC$
operator
defined in (\ref{K1}). In the Appendix we discuss this point further.

To explore the structure of the unequal-frequency Pais-Uhlenbeck model in
detail we make a further partial-fraction decomposition of the $G(E)$
propagator
in (\ref{K29}):
\begin{eqnarray}
G(E)=\frac{1}{2\omega_1(\omega_1^2-\omega_2^2)}\left(\frac{1}{E-\omega_1}-
\frac{1}{E+\omega_1}\right)-\frac{1}{2\omega_2(\omega_1^2-\omega_2^2)}\left(
\frac{1}{E-\omega_2}-\frac{1}{E+\omega_2}\right).
\label{K31}
\end{eqnarray}
In (\ref{K31}) there are two pole terms having positive coefficients and two
having negative coefficients. Whether or not these negative coefficients are
associated with negative residues depends on the way one performs the contour
integration in the complex energy plane. For the conventional Feynman
contour,
where the positive-frequency poles lie below and the negative-frequency poles
lie above the real-$E$ axis, contour integration yields the Feynman
propagator
\begin{eqnarray}
G^{\rm F}(t)&=&-\frac{1}{2\pi i}\int dE\,e^{-iEt}G(E)\nonumber\\
&=&\frac{\theta(t)}{\omega_1^2-\omega_2^2}\left(\frac{e^{-i\omega_1 t}}{2
\omega_1}-\frac{e^{-i\omega_2 t}}{2\omega_2}\right)
+\frac{\theta(-t)}{\omega_1^2-\omega_2^2}\left(\frac{e^{i\omega_1
t}}{2\omega_1}-\frac{e^{i\omega_2 t}}{2\omega_2}\right).
\label{K32}
\end{eqnarray}
In (\ref{K32}) $G^{\rm F}(t)$ describes the forward propagation of two
positive-energy particles and the backward propagation of two negative-energy
antiparticles and both of the $\omega_2$-dependent terms have negative
coefficients.

To avoid these negative coefficients we can instead choose an
unconventional contour for which the poles at $\omega_1$ and $-\omega_2$ are
taken to lie below the real $E$ axis and the poles at $-\omega_1$ and
$\omega_2$
to lie above it. For this choice, contour integration yields the
unconventional
propagator
\begin{eqnarray}
G^{\rm UNC}(t)&=&-\frac{1}{2\pi i}\int dE\,e^{-iEt}G(E)\nonumber\\
&=&\frac{\theta(t)}{\omega_1^2-\omega_2^2}\left(\frac{e^{-i\omega_1t}}{2\omega_1
}+\frac{e^{i\omega_2
t}}{2\omega_2}\right)+\frac{\theta(-t)}{\omega_1^2-\omega_2
^2}\left(\frac{e^{i\omega_1t}}{2\omega_1}+\frac{e^{-i\omega_2 t}}{2\omega_2}
\right).
\label{K33}
\end{eqnarray}
Now all of the coefficients are positive, but $G^{\rm UNC}(t)$ describes the
forward propagation of one positive-energy particle and one negative-energy
antiparticle, and also the backward propagation of one negative-energy
antiparticle and one positive-energy particle.

The $G^{\rm F}(t)$ propagator describes a system whose multiparticle energy
spectrum is bounded below [the energy eigenvalues are
$E(n_1,n_2)=(n_1+1/2)\omega_1+
(n_2+1/2)\omega_2$] but some of its poles have negative residues. The
poles of
$G^{\rm UNC}(t)$ have positive residues but the energy spectrum is unbounded
below [the energy eigenvalues are $E'(n_1,n_2)=(n_1+1/2)\omega_1-(n_2+1/2)
\omega_2$]
and thus it exhibits forward propagation of negative-energy states. The
$G^{\rm
F}(t)$ and $G^{\rm UNC}(t)$ propagators both seem to have problems, which
explains why fourth-order theories are not thought to be viable.

If we evaluate the Feynman path integral to construct the propagator, we
obtain
$G^{\rm F}(t)$ directly and do not obtain $G^{\rm UNC}(t)$. Specifically, for
the Pais-Uhlenbeck action (\ref{K26}) the path integral
\begin{equation}
G(z_i,\dot{z}_i,z_f,\dot{z}_f,T)=\int{\cal
D}(z,\dot{z})\exp\left\{i\frac{\gamma
}{2}\int_0^T dt\left[{\ddot z}^2-\left(\omega_1^2+\omega_2^2\right){\dot
z}^2+
\omega_1^2\omega_2^2z^2\right]\right\}
\label{K34}
\end{equation}
taken over all paths having fixed initial and final velocities can be
performed
analytically \cite{R22}. Taking its deep Euclidean time limit, where
$e^{-iEt}
\to e^{-E\tau}$, one finds that the low-lying energy eigenvalues are
positive:
$E=\omega_1$ and $E=\omega_2$. Also, their excitations are just as
required in
the conventional Feynman contour prescription \cite{R22}. One should not
expect
a Feynman path integral to give an energy spectrum having negative energies
because $e^{-E\tau}$ would not be finite at large $\tau$, so we do not
consider
$G^{\rm UNC}(t)$ further \cite{R23}. Our task then is to find a physically
acceptable quantum-mechanical interpretation for $G^{\rm F}(t)$.

To do this, we consider the Schr\"odinger eigenvalue problem associated
with the
Hamiltonian $H_{\rm PU}$ in (\ref{K30}). We set $p_z=-i\partial_z$, $p_x=-i
\partial_x$ and obtain
\begin{equation}
\left[-\frac{1}{2\gamma}\frac{\partial^2}{\partial x^2}-ix\frac{\partial}{
\partial z}+\frac{\gamma}{2}(\omega_1^2+\omega_2^2)x^2-\frac{\gamma}{2}
\omega_1^2\omega_2^2z^2\right]\psi_n(z,x)=E_n\psi_n(z,x),
\label{K35}
\end{equation}
which has ground-state energy $E_0=(\omega_1+\omega_2)/2$ and corresponding
eigenfunction
\begin{equation}
\psi_0(z,x)={\rm
exp}\left[\frac{\gamma}{2}(\omega_1+\omega_2)\omega_1\omega_2
z^2+i\gamma\omega_1\omega_2zx-\frac{\gamma}{2}(\omega_1+\omega_2)x^2\right].
\label{K36}
\end{equation}
As $z\to\pm\infty$, this eigenfunction diverges. Thus, in the space of
such an
eigenfunction the operator $p_z=-i\partial_z$ cannot be Dirac Hermitian.
For a
canonical commutator of the form $[z,p_z]=i$, one can only associate $p_z$
with
a differential operator $-i\partial_z$ when the commutator acts on test
functions $\psi(z)$ that are well behaved at large $z$. Since this is not the
case for the eigenfunction $\psi_0(z,x)$, we see that the operator $p_z$
is not
Hermitian on the real-$z$ axis.

The eigenfunction $\psi_0(z,x)$ vanishes exponentially rapidly for large
$|z|$
when $z$ is imaginary (or, more generally, when $z$ is confined to the two
{\it
Stokes wedges} $|{\rm Im}(z)|\geq|{\rm Re}(z)|$). To exploit this fact, we
perform an operator similarity transform of the quantum-mechanical operators
$z$ and $p_z$:
\begin{equation}
y=e^{\pi p_zz/2}ze^{-\pi p_zz/2}=-iz,\qquad q=e^{\pi p_zz/2}p_ze^{-\pi
p_zz/2}=
ip_z.
\label{K37}
\end{equation}
The commutator of the operators $y$ and $q$ still has the canonical form
$[y,q]=
i$. In terms of the operators $y$ and $q$ the Hamiltonian now takes the form
\begin{equation}
H=\frac{p^2}{2\gamma}-iqx+\frac{\gamma}{2}\left(\omega_1^2+\omega_2^2
\right)x^2+\frac{\gamma}{2}\omega_1^2\omega_2^2y^2,
\label{K38}
\end{equation}
where for notational simplicity we have replaced $p_x$ by $p$.

Since $H$ and $H_{\rm PU}$ are related by a similarity transform, they both
describe the same physics, and one can use the Hamiltonian $H$ to explore the
structure of the Pais-Uhlenbeck model \cite{R24}. In (\ref{K38}) the
operators
$p$, $x$, $q$, and $y$ are now formally Hermitian on the real-$x$ and
real-$y$
axes, but because of the $-iqx$ term $H$ has become complex and is manifestly
not Dirac Hermitian. This non-Hermiticity property was not apparent in the
original form of the Hamiltonian $H_{\rm PU}$ given in (\ref{K30}), and it is
the key to uncovering the structure of the unequal-frequency Pais-Uhlenbeck
model.

While $H$ is not Hermitian, with the $\cP$ and $\cT$ quantum-number
assignments
\begin{equation}
\begin{array}{c|cccc} &p&x&q&y\\ \hline\cP&-&-&+&+\\ \cT&-&+&+&-\\
\cP\cT&+&-&+&-\\ \end{array}
\label{K39}
\end{equation}
we see that $H$ is $\cP\cT$ symmetric. Thus, $H$ can be transformed to a
Hermitian Hamiltonian by means of the similarity transformation $\tilde
H=e^{-
\cQ/2}He^{\cQ/2}$. In Ref.~\cite{R5} the operator $\cQ$ was calculated
exactly:
\begin{equation}
\cQ=\alpha pq+\beta xy,\qquad \alpha=\frac{1}{\gamma
\omega_1\omega_2}\log\left(\frac{\omega_1+\omega_2}{\omega_1-\omega_2}\right),
\qquad\beta=\alpha\gamma^2\omega_1^2\omega_2^2
\label{K40}
\end{equation}
It leads to
\begin{eqnarray}
&&\tilde{y}=y\cosh\,\theta+i(\alpha/\beta)^{1/2}\,p\sinh\,\theta,\quad
\tilde{p}=p\cosh\,\theta-i(\beta/\alpha)^{1/2}\,y\sinh\,\theta,\nonumber \\
&&\tilde{x}=x\cosh\,\theta+i(\alpha/\beta)^{1/2}q\sinh\,\theta,\quad
\tilde{q}=q\cosh\,\theta-i(\beta/\alpha)^{1/2}\,x\sinh\,\theta,\nonumber\\
&&{\tilde
H}=\frac{{p}^2}{2\gamma}+\frac{{q}^2}{2\gamma\omega_1^2}+\frac{\gamma}
{2}\omega_1^2 x^2+\frac{\gamma}{2}\omega_1^2\omega_2^2{y}^2,
\label{K41}
\end{eqnarray}
where $\theta=(\alpha\beta)^{1/2}/2$, $\tilde{y}=e^{-\cQ/2}ye^{\cQ/2}$,
etc. The
transformed Hamiltonian $\tilde H$ is a manifestly positive-definite
operator.
Equations (\ref{K13}) -- (\ref{K15}) now follow directly and we obtain a
viable
unequal-frequency theory \cite{R25}.

\section{Fock space in the unequal-frequency case}
\label{s3}

To construct the Fock space associated with the unequal-frequency
Pais-Uhlenbeck
model, we construct the Heisenberg equations of motion associated with $H$ in
(\ref{K38}):
\begin{equation}
\dot{y}=-ix,\qquad \dot{x}=\frac{p}{\gamma},\qquad
\dot{p}=iq-\gamma(\omega_1^2+\omega_2^2)x,\qquad
\dot{q}=-\gamma\omega_1^2\omega_2^2y.
\label{K42}
\end{equation}
All four operators $y$, $x$, $p$, and $q$ obey the same fourth-order operator
differential equations,
\begin{eqnarray}
&&\frac{d^4y}{dt^4}+(\omega_1^2+\omega_2^2)\frac{d^2y}{dt^2}+\omega_1^2
\omega_2^2y=0,\qquad
\frac{d^4x}{dt^4}+(\omega_1^2+\omega_2^2)\frac{d^2x}{dt^2}+\omega_1^2
\omega_2^2x=0,\nonumber \\
&&\frac{d^4p}{dt^4}+(\omega_1^2+\omega_2^2)\frac{d^2p}{dt^2}+\omega_1^2
\omega_2^2p=0,\qquad
\frac{d^4q}{dt^4}+(\omega_1^2+\omega_2^2)\frac{d^2q}{dt^2}+\omega_1^2
\omega_2^2q=0,
\label{K43}
\end{eqnarray}
which are identical in form to the classical Pais-Uhlenbeck equation of
motion.

Despite the presence of the complex number $i$ in $H$, all the
coefficients in
the quantum-mechanical equations of motion are real. From (\ref{K43}) each of
the dynamical operators can be expanded in terms of a basis consisting of two
raising and two lowering operators having $e^{\pm i\omega_1 t}$ and $e^{\pm i
\omega_2 t}$ time dependence. From (\ref{K42}) we thus obtain
\begin{eqnarray}
&&y=-ia_1e^{-i\omega_1t}+a_2e^{-i\omega_2t}-i\hat{a}_1e^{i\omega_1t}+\hat{a}_2
e^{i\omega_2t},\nonumber\\
&&x=-i\omega_1a_1e^{-i\omega_1t}+\omega_2a_2e^{-i\omega_2t}+i\omega_1\hat{a}_1
e^{i\omega_1t}-\omega_2\hat{a}_2e^{i\omega_2t},\nonumber\\
&&p=\gamma[-\omega_1^2a_1e^{-i\omega_1t}-i\omega_2^2a_2e^{-i\omega_2t}-\omega_1
^2\hat{a}_1e^{i\omega_1t}-i\omega_2^2\hat{a}_2e^{i\omega_2t}],\nonumber\\
&&q=\gamma\omega_1\omega_2[-\omega_2a_1e^{-i\omega_1t}-i\omega_1a_2e^{-i
\omega_2t}+\omega_2\hat{a}_1e^{i\omega_1t}+i\omega_1\hat{a}_2e^{i\omega_2t}].
\label{K44}
\end{eqnarray}
The operators $\hat{a}_1$ and $\hat{a}_2$ are {\it not} the Dirac-Hermitian
adjoints of the lowering operators $a_1$ and $a_2$ because they are the
raising
operators for a non-Hermitian Hamiltonian. In terms of the raising and
lowering
operators, $H$ takes the diagonal form
\begin{equation}
H=2\gamma(\omega_1^2-\omega_2^2)\left(\omega_1^2\hat{a}_1a_1+\omega_2^2
\hat{a}_2a_2\right)+\frac{1}{2}(\omega_1+\omega_2),
\label{K45}
\end{equation}
where the operator commutation algebra is given by
\begin{eqnarray}
&&[a_1,\hat{a}_1]=[2\gamma\omega_1(\omega_1^2-\omega_2^2)]^{-1},\quad
[a_2,\hat{a}_2]=[2\gamma\omega_2(\omega_1^2-\omega_2^2)]^{-1},\nonumber\\
&&[a_1,a_2]=0,\quad[a_1,\hat{a}_2]=0,\quad[\hat{a}_1,a_2]=0,\quad
[\hat{a}_1,\hat{a}_2]=0.
\label{K46}
\end{eqnarray}

In (\ref{K45}) and (\ref{K46}) the relative signs are all positive, so these
equations define a standard two-dimensional harmonic oscillator system
\cite{R26}. The Heisenberg equations of motion for the raising and lowering
operators are
\begin{equation}
\dot{a}_1=i[H,a_1]=-i\omega_1a_1,\quad\dot{\hat{a}}_1=i\omega_1\hat{a}_1,\quad
\dot{a}_2=-i\omega_2a_2,\quad \dot{\hat{a}}_2=i\omega_2\hat{a}_2.
\label{K47}
\end{equation}
The solutions to these equations are $a_1(t)=a_1(0)e^{-i\omega_1t}$,
$\hat{a}_1
(t)=\hat{a}_1(0)e^{i\omega_1t}$, $a_2(t)=a_2(0)e^{-i\omega_2t}$,
$\hat{a}_2(t)=
\hat{a}_2(0)e^{i\omega_2t}$, just as is required of any set of raising and
lowering operators.

\section{Fock space in the equal-frequency case}
\label{s4}

The apparatus of $\cP\cT$ quantum mechanics, as developed above, is readily
implementable for the unequal-frequency Pais-Uhlenbeck model, but the
operator
$\cQ$, as given in (\ref{K40}), becomes singular in the equal-frequency limit
$\omega_1-\omega_2\to0$. Additionally, in this limit the partial-fraction
form
of $G(E)$ in (\ref{K29}) is not valid. Thus, the equal-frequency limit of the
Pais-Uhlenbeck model must  be treated separately from the unequal-frequency
case. We will see that because $\cQ$ becomes singular, the $\cP
\cT$-symmetric Hamiltonian $H$ develops a nondiagonalizable Jordan-block
structure \cite{R27}. Because this happens it is no longer possible to
construct
an equivalent Hermitian Hamiltonian $\tilde H$. Moreover, the $\cP\cT$
sector of
the equal-frequency theory becomes an independent and self-contained
realization
of quantum mechanics. An objective of this paper is to show that in the
singular
equal-frequency limit the unitarity of the Pais-Uhlenbeck theory is not lost.

To obtain the equal-frequency limit of the Pais-Uhlenbeck theory we must
find a
basis in which, unlike (\ref{K45}) and (\ref{K46}), the operators and
operator
algebra are continuous in the limit. To do this we define
\begin{equation}
\omega_1\equiv\omega+\epsilon,\qquad \omega_2\equiv\omega-\epsilon
\label{K48}
\end{equation}
and introduce the new operators \cite{R28}
\begin{eqnarray}
&&a=a_1\left(1+\frac{\epsilon}{2\omega}\right)+ia_2\left(1-\frac{\epsilon}{2
\omega}\right),\qquad b=\frac{\epsilon}{2\omega}(a_1-ia_2),\nonumber\\
&&\hat{a}=\hat{a}_1\left(1+\frac{\epsilon}{2\omega}\right)+i\hat{a}_2\left(1-
\frac{\epsilon}{2\omega}\right),\qquad \hat{b}=\frac{\epsilon}{2\omega}
(\hat{a}_1-i\hat{a}_2).
\label{K49}
\end{eqnarray}
These new operators obey the commutation algebra
\begin{equation}
[a,\hat{a}]=\lambda,\quad [a,\hat{b}]=\mu,\quad [b,\hat{a}]=\mu,\quad
[b,\hat{b}]=\lambda,\quad[a,b]=0,\quad [\hat{a},\hat{b}]=0,
\label{K50}
\end{equation}
where
\begin{equation}
\lambda=-\frac{\epsilon^2}{16\gamma(\omega^2-\epsilon^2)\omega^3},\qquad
\mu=\frac{2\omega^2-\epsilon^2}{16\gamma(\omega^2-\epsilon^2)\omega^3}.
\label{K51}
\end{equation}

In terms of these new operators the unequal-frequency position operator
$y(t)$
is
\begin{equation}
y_{\epsilon\neq0}=e^{-i\omega t}\left[-i(a-b){\rm cos}~\epsilon
t-\frac{2b\omega
}{\epsilon}{\rm sin}~\epsilon t\right]+e^{i\omega
t}\left[-i(\hat{a}-\hat{b}){
\rm cos}~\epsilon t+\frac{2\hat{b}\omega}{\epsilon}{\rm sin}~\epsilon
t\right],
\label{K52}
\end{equation}
while the unequal-frequency Hamiltonian $H$ in (\ref{K38}) is rewritten as
\begin{equation}
H_{\epsilon\neq0}=8\gamma\omega^2\epsilon^2(\hat{a}a-\hat{b}b)+8\gamma\omega^4
\left(2\hat{b}b+\hat{a}b+\hat{b}a\right)+\omega.
\label{K53}
\end{equation}

The advantage of these new operators is that $H(\epsilon)$ and $y(\epsilon)$,
and also $p(\epsilon )$, $x(\epsilon )$, and $q(\epsilon )$, are
continuous in
the $\epsilon\to0$ limit:
\begin{equation}
H_{\epsilon=0}=8\gamma\omega^4(2\hat{b}b+\hat{a}b+\hat{b}a)+\omega,
\label{K54}
\end{equation}
\begin{equation}
y_{\epsilon=0}=e^{-i\omega t}\left[-i(a-b)-2b\omega t\right]+e^{i\omega t}
\left[-i(\hat{a}-\hat{b})+2\hat{b}\omega t\right].
\label{K55}
\end{equation}
The commutation relations (\ref{K50}) are also continuous in the limit
$\epsilon
\to0$ and, together with the commutation relations with $H(\epsilon)$,
they tend
to the gaugelike form
\begin{eqnarray}
&&[a,\hat{a}]=0,\quad [a,\hat{b}]=\frac{1}{8\gamma\omega^3},\quad[b,\hat{a}]=
\frac{1}{8\gamma\omega^3},\quad[b,\hat{b}]=0,\quad[a,b]=0,\quad[\hat{a},\hat{b}
]=0,\nonumber\\
&&[H_{\epsilon=0},\hat{a}]=\omega(\hat{a}+2\hat{b}),\qquad[H_{\epsilon=0},a]=
-\omega(a+2b),\nonumber\\
&&[H_{\epsilon=0},\hat{b}]=\omega\hat{b},\qquad[H_{\epsilon=0},b]=-\omega b.
\label{K56}
\end{eqnarray}

Note that when $\epsilon\neq0$ the states $\hat{a}|\Omega\rangle$ and
$\hat{b}|
\Omega\rangle$, where $|\Omega\rangle$ is the no-particle vacuum
annihilated by
$a$ and $b$, are not eigenstates of $H_{\epsilon\neq0}$. Rather, the
action of
the unequal-frequency Hamiltonian on these states is given by
\begin{eqnarray}
H_{\epsilon\neq0}\hat{a}|\Omega\rangle&=&\frac{1}{2\omega}\left[(4\omega^2+
\epsilon^2)\hat{a}|\Omega\rangle+(4\omega^2-\epsilon^2)\hat{b}|\Omega\rangle
\right],\quad\nonumber\\
H_{\epsilon\neq0}\hat{b}|\Omega\rangle&=&\frac{1}{2\omega}\left[\epsilon^2a^{
\dagger}|\Omega\rangle+(4\omega^2-\epsilon^2)\hat{b}|\Omega\rangle\right],
\label{K57}
\end{eqnarray}
and the Hamiltonian acts in the one-particle sector as the nondiagonal matrix
\begin{eqnarray}
M_{\epsilon\neq0}=\frac{1}{2\omega}\pmatrix{4\omega^2+\epsilon^2&4\omega^2-
\epsilon^2\cr\epsilon^2&4\omega^2-\epsilon^2}.
\label{K58}
\end{eqnarray}
In the one-particle sector the Hamiltonian has two eigenstates
\begin{eqnarray}
H_{\epsilon\neq0}|2\omega\pm\epsilon\rangle=(2\omega\pm\epsilon)|2\omega\pm
\epsilon\rangle,
\label{K59}
\end{eqnarray}
where
\begin{equation}
|2\omega\pm\epsilon\rangle=\left[\pm\epsilon\hat{a}+(2\omega\mp\epsilon)\hat{b}
\right]|\Omega\rangle.
\label{K60}
\end{equation}
A similarity transformation diagonalizes $M_{\epsilon\neq0}$:
\begin{eqnarray}
S^{-1}\left(\frac{1}{2\omega}\right)\pmatrix{4\omega^2+\epsilon^2&4\omega^2-
\epsilon^2\cr\epsilon^2&4\omega^2-\epsilon^2}S
=\pmatrix{2\omega+\epsilon&0\cr 0&2\omega-\epsilon},
\label{K61}
\end{eqnarray}
where
\begin{eqnarray}
S&=&\frac{1}{2\epsilon\omega^{1/2}(2\omega+\epsilon)^{1/2}}\pmatrix{2\omega+
\epsilon&-(4\omega^2-\epsilon^2)\epsilon\cr\epsilon&(2\omega+\epsilon)\epsilon^2
},\nonumber\\
S^{-1}&=&\frac{1}{2\epsilon\omega^{1/2}(2\omega+\epsilon)^{1/2}}\pmatrix{(2
\omega+\epsilon)\epsilon^2&(4\omega^2-\epsilon^2)\epsilon\cr-\epsilon&2\omega
+\epsilon}.
\label{K62}
\end{eqnarray}

Now let us examine the limit $\epsilon\to0$. From (\ref{K60}) we can see
that as
$\epsilon\to0$ the two eigenstates $|2\omega\pm\epsilon\rangle$ collapse onto
one state $\hat{b}|\Omega\rangle$. Thus, $H(\epsilon)$ loses a
one-particle eigenstate in
this limit. This is to be expected because the commutator
$[H_{\epsilon=0},\hat{
a}]$ contains components parallel to both $\hat {a}$ and $\hat{b}$. To
understand why an eigenstate has disappeared, we note that as $\epsilon \to0$
the matrix $M(\epsilon)$ in (\ref{K58}) takes the upper-triangular
Jordan-block
form
\begin{eqnarray}
M_{\epsilon=0}=2\omega\pmatrix{1&1\cr 0&1}.
\label{K63}
\end{eqnarray}
The matrix $M_{\epsilon=0}$ possesses two eigenvalues, both equal to
$2\omega$,
but it has only one eigenstate because the eigenvector condition
\begin{eqnarray}
\pmatrix{1&1\cr 0&1}\pmatrix{c \cr d}=\pmatrix{c+d\cr d}=\pmatrix{c\cr d}
\label{K64}
\end{eqnarray}
permits only one solution, namely that with $d=0$. Despite not being Dirac
Hermitian and despite having lost an eigenstate, $H_{\epsilon=0}$
continues to
be $\cP\cT$ symmetric under the transformations (\ref{K39}), and thus its
eigenvalues remain real. Furthermore, as $\epsilon\to0$, $S$ and $S^{-1}$ in
(\ref{K62}) both become singular, so it is not possible to diagonalize $M_{
\epsilon=0}$.

This same pattern occurs for the higher excited states of $H$. (For
example, one
can readily show that the three two-particle states collapse onto one common
eigenstate as $\epsilon\to0$.) To discuss the $\epsilon\to0$ limit for
arbitrary
multiparticle states, we return to the operator $\cQ$ in (\ref{K40}) and note
that at $t=0$ we can use (\ref{K44}) to write
\begin{eqnarray}
pq&+&\gamma^2\omega_1^2\omega_2^2xy\nonumber\\
&=&i\gamma^2\omega_1\omega_2(\omega_1^2-\omega_2^2)\left[(\omega_1-\omega_2)(a_1
a_2-\hat{a}_1\hat{a}_2)+(\omega_1+\omega_2)(\hat{a}_1a_2-\hat{a}_2a_1)\right].
\label{K65}
\end{eqnarray}
In the $\epsilon\to0$ limit we thus obtain
\begin{equation}
pq+\gamma^2\omega_1^2\omega_2^2xy\to8\gamma^2\omega^5(\hat{b}^2-b^2+\hat{b}a-
\hat{a}b).
\label{K66}
\end{equation}
The quantity $pq+\gamma^2\omega_1^2\omega_2^2xy$ is well-behaved as
$\epsilon\to
0$, and the singularity acquired by the coefficient $\alpha$ in (\ref{K40})
cannot be canceled. The similarity transform $e^{-\cQ/2}$ in (\ref{K3}) also
becomes singular.

Since the operator $\cQ$ creates two-particle pairs in (\ref{K65}) and
(\ref{K66}), $e^{-\cQ/2}$ is a singular operator in every multiparticle
sector
of $H$, with each such sector developing Jordan-block structure in the limit.
The equal-frequency Pais-Uhlenbeck model possesses no equivalent
Dirac-Hermitian
counterpart; as we show in the next sections, it constitutes a self-contained
realization of quantum mechanics that exists in its own right.

\section{Eigenfunctions of the Unequal-Frequency Theory}
\label{s5}

The outstanding property of a Hamiltonian in Jordan-block form is that its
eigenstates are incomplete. One may ask, where do the other eigenstates go in
the equal-frequency limit, and how can one formulate quantum mechanics if the
space of energy eigenstates is incomplete? To answer these questions for the
Pais-Uhlenbeck model, we must construct the eigenfunctions of the
unequal-frequency theory, where there are no Jordan-block structures, and we
must then track what happens to the eigenfunctions in the equal-frequency
limit.

Noting that the coordinate-space representation of the Hamiltonian
(\ref{K38})
is not symmetric because it has a term that acts like $x\partial_y$, we will
need to distinguish between a right Hamiltonian, which operates to the right,
and a left Hamiltonian, which operates to the left. When acting to the
right on
well-behaved states, the commutator $[y,q]=i$ is realized by setting $q=-i
\partial_y$, but when acting to the left it is necessary to use $q=+i
\partial_y$. We thus obtain the right and left Schr\"odinger equations
\begin{equation}
i\frac{\partial \psi^{\rm R}_n(x,y,t)}{\partial
t}=\left[-\frac{1}{2\gamma}\frac
{\partial^2}{\partial x^2}-x\frac{\partial}{\partial y}+\frac{\gamma}{2}(
\omega_1^2+\omega_2^2)x^2+\frac{\gamma}{2}\omega_1^2\omega_2^2y^2\right]
\psi_n^{\rm R}(x,y,t),
\label{K67}
\end{equation}
\begin{equation}
-i\frac{\partial \psi^{\rm L}_n(x,y,t)}{\partial t}=\left[-\frac{1}{2\gamma}
\frac{\partial^2}{\partial x^2}+x\frac{\partial}{\partial
y}+\frac{\gamma}{2}(
\omega_1^2+\omega_2^2)x^2+\frac{\gamma}{2}\omega_1^2\omega_2^2y^2\right]
\psi_n^{\rm L}(x,y,t).
\label{K68}
\end{equation}

The ground state, whose energy is $E_0=(\omega_1+\omega_2)/2$, has right and
left eigenfunctions
\begin{equation}
\psi^{\rm R}_0(x,y,t)=\exp\left[-\frac{\gamma}{2}(\omega_1+\omega_2)\omega_1
\omega_2y^2-\gamma\omega_1\omega_2yx-\frac{\gamma}{2}(\omega_1+\omega_2)x^2-
\frac{i}{2}(\omega_1+\omega_2)t\right],
\label{K69}
\end{equation}
\begin{equation}
\psi^{\rm L}_0(x,y,t)=\exp\left[-\frac{\gamma}{2}(\omega_1+\omega_2)\omega_1
\omega_2y^2+\gamma\omega_1\omega_2yx-\frac{\gamma}{2}(\omega_1+\omega_2)x^2+
\frac{i}{2}(\omega_1+\omega_2)t\right].
\label{K70}
\end{equation}
For this ground state we can define a normalization integral of the form
\begin{eqnarray}
N_0&=&\int dx\,dy\,\psi^{\rm L}_0(x,y,t)\psi^{\rm R}_0(x,y,t)\nonumber\\
&=&\int dx\,dy\exp\left[-\gamma(\omega_1+\omega_2)\omega_1\omega_2y^2-\gamma(
\omega_1+\omega_2)x^2\right]=\frac{\pi}{\gamma(\omega_1+\omega_2)(\omega_1\omega
_2)^{1/2}},
\label{K71}
\end{eqnarray}
with $N_0$ being time independent, finite, and real.

The two one-particle states with energies $E_1=E_0+\omega_1$,
$E_2=E_0+\omega_2$
have eigenfunctions
\begin{eqnarray}
&&\psi^{\rm R}_1(x,y,t)=(x+\omega_2y)\psi_0^{\rm
R}(x,y,t)e^{-i\omega_1t},\quad
\psi^{\rm L}_1(x,y,t)=(x-\omega_2y)\psi_0^{\rm
L}(x,y,t)e^{i\omega_1t},\nonumber
\\
&&\psi^{\rm R}_2(x,y,t)=(x+\omega_1y)\psi_0^{\rm
R}(x,y,t)e^{-i\omega_2t},\quad
\psi^{\rm L}_2(x,y,t)=(x-\omega_1y)\psi_0^{\rm L}(x,y,t)e^{i\omega_2t},
\label{K72}
\end{eqnarray}
whose normalization integrals are
\begin{equation}
N_1=\frac{\pi(\omega_1-\omega_2)}{2\gamma^{2}(\omega_1+\omega_2)^{2}\omega_1^{
3/2}\omega_2^{1/2}},\qquad N_2 =-\frac{\pi(\omega_1-\omega_2)}{2\gamma^{2}(
\omega_1+\omega_2)^{2}\omega_1^{1/2}\omega_2^{3/2}}.
\label{K73}
\end{equation}
Again, these normalizations are time independent, finite, and real, but
$N_2$ is negative, a crucial issue that we return to and resolve below.

Analogously, the three two-particle states with energies $E_3=E_0+2\omega_1$,
$E_4=E_0+\omega_1+\omega_2$, and $E_5=E_0+2\omega_2$ have right and left
eigenfunctions
\begin{eqnarray}
&&\psi_3^{\rm
R}(x,y,t)=\left[(x+\omega_2y)^2-\frac{1}{2\gamma\omega_1}\right]
\psi_0^{\rm R}(x,y,t)e^{-2i\omega_1t},\nonumber\\ &&\psi_4^{\rm
R}(x,y,t)=\left[(x+\omega_1y)(x+\omega_2y)-\frac{1}{\gamma(\omega
_1+\omega_2)}\right]\psi_0^{\rm
R}(x,y,t)e^{-i\omega_1t-i\omega_2t},\nonumber\\
&&\psi_5^{\rm
R}(x,y,t)=\left[(x+\omega_1y)^2-\frac{1}{2\gamma\omega_2}\right]
\psi_0^{\rm R}(x,y,t)e^{-2i\omega_2t},
\label{K74}
\end{eqnarray}
and
\begin{eqnarray}
&&\psi^{\rm
L}_3(x,y,t)=\left[(x-\omega_2y)^2-\frac{1}{2\gamma\omega_1}\right]
\psi^{\rm L}_0(x,y,t)e^{2i\omega_1t},\nonumber\\ &&\psi^{\rm
L}_4(x,y,t)=\left[(x-\omega_1y)(x-\omega_2y)-\frac{1}{\gamma(\omega
_1+\omega_2)}\right]\psi^{\rm
L}_0(x,y,t)e^{i\omega_1t+i\omega_2t},\nonumber\\
&&\psi^{\rm
L}_5(x,y,t)=\left[(x-\omega_1y)^2-\frac{1}{2\gamma\omega_2}\right]
\psi^{\rm L}_0(x,y,t)e^{2i\omega_2t}.
\label{K75}
\end{eqnarray}
The normalization integrals are time independent:
\begin{eqnarray}
&&N_3=\frac{\pi(\omega_1-\omega_2)^2}{2\gamma^3(\omega_1+\omega_2)^3(\omega_1
\omega_2)^{1/2}\omega_1^2},\nonumber\\
&&N_4=-\frac{\pi(\omega_1-\omega_2)^2}{4\gamma^3(\omega_1+\omega_2)^3(\omega_1
\omega_2)^{1/2}\omega_1\omega_2},\nonumber\\
&&N_5=\frac{\pi(\omega_1-\omega_2)^2}{2\gamma^3(\omega_1+\omega_2)^3(\omega_1
\omega_2)^{1/2}\omega_2^2}.
\label{K76}
\end{eqnarray}
Note that $N_4$ is negative.

To relate these normalization integrals to the Hilbert-space norms of the
eigenstates of $H$, we recall that the energy eigenstates of the
Hamiltonian are
eigenstates of both the $\cP\cT$ operator and of the $\cC=e^{\cQ}\cP$
operator
introduced in Sec.~\ref{s1}. The general procedure for constructing such
states
for symmetric Hamiltonians is given in Ref.~\cite{R3}, and we adapt it
here for
the nonsymmetric case. Since the $\cP\cT$ operator is antilinear, in
general its
eigenstates have eigenvalue $e^{i\alpha}$, where $\alpha$ is a real phase
that
depends on the eigenstate. Multiplying each eigenfunction by $e^{i\alpha/2}$
gives a new eigenfunction that is still an eigenstate of the Hamiltonian, but
now it has $\cP\cT$ eigenvalue equal to one. Because the operator $\cC$ obeys
the algebraic conditions in (\ref{K1}), its eigenstates are also
eigenstates of
$\cP\cT$ and of $H$ and all of its eigenvalues $\cC_n$ are $\pm1$.
Consequently,
each energy eigenstate is an eigenstate of $\cC\cP\cT$ with eigenvalue
$\cC_n=
\pm1$.

When the Hamiltonian $H$ is symmetric, the completeness and normalization
conditions have the form \cite{R3}
\begin{eqnarray}
&&\sum_ne^{iE_nt}[\cC\cP\cT\psi_n(x',y',t=0)]\psi_n(x,y,t=0)e^{-iE_nt}=\delta
(x-x')\delta(y-y'),\nonumber\\
&&\int dx\,dy\,e^{iE_nt}[\cC\cP\cT\psi_n(x,y,t=0)]\psi_m(x,y,t=0)e^{-iE_nt}=
\delta_{m,n},
\label{K77}
\end{eqnarray}
where the summation is taken over all the eigenstates of the Hamiltonian. For
the Pais-Uhlenbeck oscillator, $H$ is not symmetric and thus we must
distinguish
between right and left wave functions. We identify $\psi_n^{\rm
L}(x',y',t=0)=
\cC\cP\cT\psi_n^{\rm R}(x',y',t=0)$ and then assign the $\cC_n$
eigenvalues as
follows: We take $\cC_0=1$ for the ground state $\psi_0(x,y,t)$, $\cC_1=1$
for
the one-particle state $\psi_1(x,y,t)$ and $\cC_2=-1$ for the one-particle
state
$\psi_2(x,y,t)$. The three two-particle states in (\ref{K74}) then acquire
eigenvalues $\cC_3=1$, $\cC_4=-1$, $\cC_5=1$.

The alternations in sign of the $\cC_n$ parallel the alternations in sign
of the
normalization integrals given in (\ref{K71}), (\ref{K73}), and
(\ref{K76}), with
the same pattern repeating for the higher multiparticle states; that is,
negative signs occur when the number of $\omega_2$ quanta is odd. Thus,
for any
eigenstate the sign of $\cC_n$ is precisely the same as that of its
normalization integral $N_n=\int dx\,dy\,\psi^{\rm L}_n(x,y,t)\psi^{\rm
R}_n(x,y
,t)$. Consequently, from (\ref{K77}) the correct completeness relation and
normalization conditions for the states of the theory are
\begin{equation}
\sum_n\psi_n^{\rm L}(x',y',t)\frac{\cC_n}{|N_n|}\psi_n^{\rm R}(x,y,t)=\delta
(x-x')\delta(y-y'),
\label{K78}
\end{equation}
\begin{equation}
\int dx\,dy\,\psi_n^{\rm L}(x,y,t)\frac{\cC_n}{|N_n|}\psi_m^{\rm R}(x,y,t)=
\delta_{m,n}.
\label{K79}
\end{equation}
Both this $\cC\cP\cT$ norm and the Fock space norm $\langle
n|e^{-\cQ}|m\rangle=
\delta_{m,n}$ in (\ref{K13}) are positive. Thus, there are no negative-norm
states in the unequal-frequency Fock space and the unequal-frequency
Pais-Uhlenbeck model is unitary.

Because the unequal-frequency energy eigenstates are complete, we can
expand an
arbitrary wave function in terms of them as $\psi^{\rm R}(x,y,t)=\sum_n
a_n\psi^
{\rm R}_n(x,y,t)$ and $\psi^{\rm L}(x,y,t)=\sum_n a_n^*\psi^{\rm
L}_n(x,y,t)$.
Then, since the energy eigenstates form an orthonormal basis with real energy
eigenvalues, it follows that the norm $\int dx\,dy\,\psi^{\rm L}(x,y,t)\cC
\psi^{\rm R}(x,y,t)$ is both time independent and real. Therefore,
probability
is preserved and the unequal-frequency theory is unitary under time
evolution.

Given the norm in (\ref{K79}) and writing the norm in (\ref{K13}) in the form
$\langle n|e^{-\cQ}|m\rangle=\int dx\,dy\,\langle n|e^{-\cQ}|x,y\rangle
\langle x,y|m\rangle$, we make the identifications
\begin{equation}
\frac{\cC_n}{|N_n|^{1/2}}\psi^{\rm L}_n(x,y,t) =\langle
n|e^{-\cQ}|x,y,t\rangle,
\qquad\frac{1}{|N_n|^{1/2}}\psi^{\rm R}_n(x,y,t)=\langle x,y,t|n\rangle.
\label{K80}
\end{equation}
Then, by using the relations in (\ref{K80}) and substituting
$H=\sum_n|n\rangle
E_n\langle n|e^{-\cQ}$ into the expression $G(x,y,x',y',t)=\langle
x,y,t=0|e^{-i
Ht}|x',y',t=0\rangle$ for the propagator, we obtain
\begin{equation}
G(x,y,x',y',t)=\sum_n\psi^{\rm
R}_n(x,y,t=0)\frac{\cC_n}{|N_n|}e^{-iE_nt}\psi^{
\rm L}_n(x',y',t=0).
\label{K81}
\end{equation}
Thus, the negative sign of $\cC_2$ accounts for the negative sign of the
$\omega_2$-dependent term in the propagators of (\ref{K29}) and (\ref{K32}).
This shows that the good convergence associated with fourth-order propagators
can be achieved in a Hilbert space whose norms are all positive and need
not be
in conflict with the requirement of unitarity.

\section{Eigenfunctions of the Equal-Frequency Theory}
\label{s6}

In the equal-frequency limit the Hamiltonian (\ref{K38}) takes the form
\begin{equation}
H=\frac{p^2}{2\gamma}-iqx+\gamma\omega^2x^2+\frac{\gamma}{2}\omega^4y^2.
\label{K82}
\end{equation}
Its ground state has energy $E_0=\omega$, left and right eigenfunctions
\begin{eqnarray}
\hat{\psi}_0^{\rm R}(x,y,t)&=& \exp\left[-\gamma\omega^3y^2
-\gamma\omega^2yx -\gamma\omega x^2-i\omega t\right],\nonumber\\
\hat{\psi}^{\rm L}_0(x,y,t)&= &\exp\left[-\gamma\omega^3y^2
+\gamma\omega^2yx -\gamma\omega x^2+i\omega t\right],
\label{K83}
\end{eqnarray}
and a normalization integral
\begin{equation}
\hat{N}_0=\int dx\,dy\,\hat{\psi}^{\rm L}_0(x,y,t)\hat{\psi}^{\rm R}_0(x,y,t)
=\frac{\pi}{2\gamma\omega^2},
\label{K84}
\end{equation}
which is time independent, finite, and real.

The equal-frequency theory differs from the unequal-frequency theory in that
there is only a single one-particle eigenstate instead of two one-particle
eigenstates. The energy of this state is $E_1=2\omega$ and its
eigenfunction is
\begin{equation}
\hat{\psi}^{\rm R}_1(x,y,t)=(x+\omega y)\hat{\psi}^{\rm
R}_0(x,y,t)e^{-i\omega
t},\qquad\hat{\psi}^{\rm L}_1(x,y,t)=(x-\omega y)\hat{\psi}^{\rm L}_0(x,y,t)
e^{i\omega t}.
\label{K85}
\end{equation}
In the equal-frequency limit the two unequal-frequency eigenstates
$\psi_1^{\rm
R}(x,y,t)$ and $\psi_2^{\rm R}(x,y,t)$ of (\ref{K72}) collapse onto one state
$\hat{\psi}_1^{\rm R}(x,y,t)=[\psi_1^{\rm R}(x,y,t)+\psi_2^{\rm R}(x,y,t)]/2$
(and likewise for the left eigenfunction).

The disappearance of eigenstates in the equal-frequency limit is generic.
This
same collapse of eigenstates occurs for the higher excited eigenstates,
with the
three unequal-frequency two-particle eigenfunctions in (\ref{K74}) and
(\ref{K75}) collapsing onto a single equal-frequency second excited
eigenstate
having energy $E_2=3\omega$:
\begin{eqnarray}
\hat{\psi}^{\rm R}_2(x,y,t)&=&\left[(x+\omega y)^2-\frac{1}{2\gamma\omega}
\right]\hat{\psi}^{\rm R}_0(x,y,t) e^{-2i\omega t},\nonumber\\
\hat{\psi}^{\rm L}_2(x,y,t)&=&\left[(x-\omega y)^2-\frac{1}{2\gamma\omega}
\right]\hat{\psi}^{\rm L}_0(x,y,t)e^{2i\omega t}.
\label{K86}
\end{eqnarray}

Even though the normalization integral $\hat{N}_0$ for the equal-frequency
ground state is positive, for the equal-frequency first excited state we find
that
\begin{equation}
\hat{N}_1=\int dx\,dy\,\hat{\psi}^{\rm L}_1(x,y,t)\hat{\psi}_1^{\rm
R}(x,y,t)=
\int dx\,dy\,(x^2-\omega^2y^2)e^{-2\gamma\omega^3y^2-2\gamma\omega x^2}=0.
\label{K87}
\end{equation}
The norm of this state vanishes because the unequal-frequency normalization
integrals $N_1$ and $N_2$ in (\ref{K73}) both vanish in the equal-frequency
limit, and the unequal-frequency states of (\ref{K72}) are orthogonal before
the limit is taken. Thus, $\hat{\psi}_1^{\rm R}(x,y,t)$ ends up being
orthogonal
to itself, that is, to $\hat{\psi}_1^{\rm L}(x,y,t)$. This same situation
repeats for the higher excited states with all of the two-particle norms of
(\ref{K76}) collapsing onto $\hat{N}_2=0$.

The emergence of zero-norm states is characteristic of Hamiltonians having
Jordan-block structure. For the two-dimensional Jordan-block matrix
(\ref{K64}),
the right eigenvector is given by the column $(1,0)$, while the left
eigenvector
is given by the row $(0,1)$ rather than by the row $(1,0)$. Thus, the norm,
which is the product of the left and right eigenvectors, is zero. Yet,
despite
the presence of zero norms, we will show below that probabilities in the
equal-frequency theory are still nonzero.

The energy eigenstates that disappear in the equal-frequency limit are
replaced
by an equal number of nonstationary states. To demonstrate this
phenomenon, we
form linear combinations of the unequal-frequency eigenfunctions
$\psi_1^{\rm R}
(x,y,t)$ and $\psi_2^{\rm R}(x,y,t)$ with coefficients that behave like $1/(
\omega_1-\omega_2)$ and then track the limit. To do this, we extract the
terms
in $\psi_1^{\rm R}(x,y,t)$ and $\psi_2^{\rm R}(x,y,t)$ that are linear in
$\omega_1-\omega_2$. This yields the nonstationary equal-frequency state
\begin{eqnarray}
\hat{\psi}_{1a}^{\rm R}(x,y,t)&=&\lim_{\epsilon\to0}\frac{\psi_2^{\rm
R}(x,y,t)- \psi_1^{\rm R}(x,y,t)}{2\epsilon}=\left[(x+\omega
y)it+y\right]\hat{\psi}_0^{\rm R}(x,y,t)e^{-i\omega t},\nonumber\\
\hat{\psi}_{1a}^{\rm L}(x,y,t)&=&\lim_{\epsilon\to0}\frac{\psi_2^{\rm
L}(x,y,t)-
\psi_1^{\rm L}(x,y,t)}{2\epsilon}=\left[-(x-\omega
y)it-y\right]\hat{\psi}_0^{
\rm L}(x,y,t)e^{i\omega t}.
\label{K88}
\end{eqnarray}
Since this state is not stationary, it is not an eigenstate of $H$.
However, it
does satisfy the {\it time-dependent} equal-frequency Schr\"odinger equation:
\begin{eqnarray}
i\frac{\partial}{\partial t}\hat{\psi}^{\rm
R}(x,y,t)=\left(-\frac{1}{2\gamma}
\frac{\partial^2}{\partial x^2}-x\frac{\partial}{\partial
y}+\gamma\omega^2x^2+
\frac{\gamma}{2}\omega^4y^2\right)\hat{\psi}^{\rm R}(x,y,t),\nonumber\\
-i\frac{\partial}{\partial t}\hat{\psi}^{\rm
L}(x,y,t)=\left(-\frac{1}{2\gamma}
\frac{\partial^2}{\partial x^2}+x\frac{\partial}{\partial
y}+\gamma\omega^2x^2+
\frac{\gamma}{2}\omega^4y^2\right)\hat{\psi}^{\rm L}(x,y,t).
\label{K89}
\end{eqnarray}
Similarly, on expanding the unequal-frequency two-particle states of
(\ref{K74})
to order $(\omega_1-\omega_2)^2$, one obtains [in addition to
$\hat{\psi}^{\rm R
}_2(x,y,t)$] the two states
\begin{eqnarray}
\hat{\psi}^{\rm R}_{2a}(x,y,t)&=&\lim_{\epsilon\to0}\frac{\psi_5^{\rm
R}(x,y,t)-
\psi_3^{\rm R}(x,y,t)}{2\epsilon}\nonumber\\ &=&\left[\left((x+\omega
y)^2-\frac{1}{2\gamma\omega}\right)2it+2xy+2\omega y^2-
\frac{1}{2\gamma\omega^2}\right]\hat{\psi}^{\rm R}_0(x,y,t)e^{-2i\omega t},
\label{K90}
\end{eqnarray}
and
\begin{eqnarray}
\hat{\psi}_{2b}^{\rm R}(x,y,t)&=&\lim_{\epsilon\to0}\frac{2\psi_4^{\rm
R}(x,y,t)
-\psi_3^{\rm R}(x,y,t)-\psi_5^{\rm R}(x,y,t)}{2\epsilon^2}\nonumber\\
&=&\left[\left((x+\omega y)^2-\frac{1}{2\gamma \omega}\right)2t^2-\left(2xy+2
\omega y^2-\frac{1}{2\gamma\omega^2}\right)2it-2y^2+\frac{1}{2\gamma\omega^3}
\right]\nonumber\\ &&\times\hat{\psi}_0^{\rm R}(x,y,t)e^{-2i\omega t}.
\label{K91}
\end{eqnarray}
Both of these states satisfy the equal-frequency time-dependent Schr\"odinger
equation even though neither state is stationary.

The picture is now clear. The unequal-frequency states $\psi_1^{\rm
R}(x,y,t)$
and $\psi_2^{\rm R}(x,y,t)$ are energy eigenstates, but linear
combinations of
them are not because the states are not degenerate. However, such linear
combinations are still solutions to the unequal-frequency time-dependent
Schr\"odinger equation. When $\epsilon\to0$, we obtain two new wave functions
$\hat{\psi}_1^{\rm R}(x,y,t)$ and $\hat{\psi}_{1a}^{\rm R}(x,y,t)$, which are
solutions to the equal-frequency time-dependent Schr\"odinger equation. One
state is stationary, so it also solves the equal-frequency time-independent
Schr\"odinger equation and thus it is an energy eigenstate, while the other
state is not. The same pattern repeats for the higher excited states. The
equal-frequency Hamiltonian is a Jordan-block matrix that has fewer
eigenstates
than eigenvalues, and the form of the unequal-frequency Hamiltonian becomes
Jordan block as $\epsilon\to0$. The counting of wave functions is
continuous in
the limit, and it is just the counting of stationary states that is not
continuous. The nonstationary states replace the missing stationary states in
the limit \cite{R29}.

Instead of counting energy eigenstates, if we count Fock states, we see
that the
unequal-frequency Fock space built from the raising operators $\hat{a}_1$ and
$\hat{a}_2$ in (\ref{K46}) and the equal-frequency Fock space built from the
raising operators $\hat{a}$ and $\hat{b}$ in (\ref{K56}) both have the
dimensionality of a two-dimensional harmonic oscillator. The eigenspectrum of
the equal-frequency Hamiltonian $H_{\epsilon=0}$ is like that of a
one-dimensional harmonic oscillator. However, the counting of states in
the full
Fock space is continuous in the limit $\epsilon\to0$.

Given the form of the Schr\"odinger equation in (\ref{K89}), any pair of its
solutions $\hat{\psi}_{A}^{\rm R}(x,y,t)$ and $\hat{\psi}_{B}^{\rm L}(x,y,t)$
obey
\begin{equation}
i\frac{\partial}{\partial t}\int dx\,dy\,\hat{\psi}_{B}^{\rm L}(x,y,t)
\hat{\psi}_{A}^{\rm R}(x,y,t)=-\int dx\,dy\,x\frac{\partial}{\partial
y}\left[
\hat{\psi}_{B}^{\rm L}(x,y,t)\hat{\psi}_{A}^{\rm R}(x,y,t)\right].
\label{K92}
\end{equation}
Because the product $\hat{\psi}_{0}^{\rm L}(x,y,t)\hat{\psi}_{0}^{\rm
R}(x,y,t)=
\exp(-2\gamma\omega^3y^2-2\gamma\omega x^2)$ is exponentially suppressed, the
overlap integrals for any pair of multiparticle eigenfunctions are time
independent:
\begin{equation}
i\frac{\partial}{\partial t}\int dx\,dy\,\hat{\psi}_{B}^{\rm L}(x,y,t)
\hat{\psi}_{A}^{\rm R}(x,y,t)=0,
\label{K93}
\end{equation}
where the indices $A$ and $B$ are $0$, $1$, $1a$, $2$, $2a$, $2b$, and so on.
(These eigenfunctions have the form of polynomials in $x$ and $y$
multiplied by
the ground-state eigenfunction.) Because $H$ is not Dirac Hermitian,
(\ref{K93})
has the generic time-independent form in (\ref{K24}).

For modes that are energy eigenstates, (\ref{K93}) reduces to
\begin{equation}
(E_A-E_B)\int dx\,dy\,\hat{\psi}_B^{\rm L}(x,y,t)\hat{\psi}_A^{\rm
R}(x,y,t)=0,
\label{K94}
\end{equation}
so modes having unequal-energy eigenvalues are orthogonal as usual.
When $A$ is an eigenmode, the diagonal integrals $\int
dx\,dy\,\hat{\psi}_A^{\rm
L}(x,y,t)\hat{\psi}_A^{\rm R}(x,y,t)$ all vanish except for the ground-state
eigenmode because all eigenstates other than the ground state have zero norm.
The overlap integrals of the nonstationary wave functions, either with each
other or with the stationary solutions, need not vanish because (\ref{K93})
requires that such overlaps be time independent, and not that they vanish.
However, the equal-frequency wave functions are constructed as limits of
linear
combinations of unequal-frequency modes, so the orthogonality of
multiparticle
eigenstates with different numbers of particles in the unequal-frequency
theory
translates into the orthogonality of the associated states in the
equal-frequency case. Thus, even though not all of the equal-frequency theory
states are stationary, both of the $1$ and $1a$ states are orthogonal to
all of
the $2$, $2a$, $2b$ states, and so on.

Within any given multiparticle sector the overlaps need not all be zero.
For the
typical one-particle sector, for instance, we evaluate the overlap integrals
directly and find that
\begin{eqnarray}
&&\int dx\,dy\,\hat{\psi}^{\rm L}_{1}(x,y,t)\hat{\psi}^{\rm R}_{1}(x,y,t)=0,
\nonumber\\
&&\int dx\,dy\,\hat{\psi}^{\rm L}_{1}(x,y,t)\hat{\psi}^{\rm
R}_{1a}(x,y,t)=\int
dx\,dy\,\hat{\psi}^{\rm L}_{1a}(x,y,t)\hat{\psi}^{\rm
R}_{1}(x,y,t)=-\frac{\pi}
{8\gamma^2\omega^4},\nonumber\\
&&\int dx\,dy\,\hat{\psi}^{\rm L}_{1a}(x,y,t)\hat{\psi}^{\rm R}_{1a}(x,y,t)=-
\frac{\pi}{8\gamma^2\omega^5}.
\label{K95}
\end{eqnarray}
This confirms that these overlaps are indeed time independent. The overlap
integral $\int dx\,dy\,\hat{\psi}^{\rm L}_{1}(x,y,t)\hat{\psi}^{\rm
R}_{1a}(x,y,
t)$, for example, is time independent because the coefficient of the term
that
is linear in $t$ is just the zero norm $\int dx\,dy\,\hat{\psi}^{\rm
L}_{1}(x,y,
t)\hat{\psi}^{\rm R}_{1}(x,y,t)$.

An alternative way to derive the relations in (\ref{K95}) is to take the
$\epsilon\to0$ limit of the normalization integrals $N_1$ and $N_2$ in
(\ref{K73}). To order $\epsilon$ this yields
\begin{eqnarray}
&&\int dx\,dy\,[\hat{\psi}_1^{\rm L}(x,y)\hat{\psi}^{\rm R}_1(x,y)
-\epsilon\hat{\psi}^{\rm L}_{1a}(x,y)\hat{\psi}^{\rm R}_1(x,y)
-\epsilon\hat{\psi}^{\rm L}_1(x,y)\hat{\psi}^{\rm R}_{1a}( x,y)]=\frac{\pi
\epsilon}{4\gamma^2\omega^4},\nonumber\\
&&\int dx\,dy\,[\hat{\psi}_1^{\rm L}(x,y)\hat{\psi}^{\rm R}_1(x,y)
+\epsilon\hat{\psi}^{\rm L}_{1a}(x,y)\hat{\psi}^{\rm R}_1(x,y)
+\epsilon\hat{\psi}^{\rm L}_1(x,y)\hat{\psi}^{\rm R}_{1a}( x,y)]
=-\frac{\pi\epsilon}{4\gamma^2\omega^4},
\label{K96}
\end{eqnarray}
from which the relevant relations in (\ref{K95}) follow. This procedure
implies
that in taking the $\epsilon\to0$ limit of the time-independent normalization
integrals of the unequal-frequency theory, we get time-independent
expressions,
some of which are nonzero, because as (\ref{K73}) and (\ref{K76}) show, the
unequal-frequency-theory normalization integrals only vanish as powers of
$\epsilon$.

The nonstationary wave functions are not energy eigenstates, but at $t=0$ the
various multiparticle wave functions $\hat{\psi}^{\rm R}_{n}(x,y,t)$ and
$\hat{
\psi}^{\rm R}_{n,\alpha}(x,y,t)$ ($n=0,1,2,...$, $\alpha=a,b,...$) contain
polynomials in $x$ and $y$ of degree $n$, with just enough freedom to
construct
any arbitrary polynomial function of $x$ and $y$. Hence any initial wave
function $\hat{\psi}^{\rm R}(x,y,t=0)$ of $x$ and $y$ can be expanded in
terms of
a complete basis of polynomial wave functions: $\hat{\psi}^{\rm
R}(x,y,t=0)=\sum_{n}
a_{n}\hat{\psi}^{\rm
R}_{n}(x,y,t=0)+\sum_{n,\alpha}a_{n,\alpha}\hat{\psi}^{\rm
R}_{n,\alpha}(x,y,t=0)$.

The left wave functions have analogous expansions, with relations such as
those
in (\ref{K95}) implying that the quantity $\hat{N}(\hat{\psi},t=0)=\int
dx\,dy\,
\hat{\psi}^{\rm L}(x,y,t=0)\hat{\psi}^{\rm R}(x,y,t=0)$ is nonzero. Thus,
we can use the
nonstationary solutions to construct any initial state whose initial
probability
$\hat{N}(\hat{\psi},t=0)$ is nonzero despite the presence of zero-norm energy
eigenstates.

Each of the $\hat{\psi}^{\rm R}_{n}(x,y,t=0)$ and $\hat{\psi}^{\rm
R}_{n,\alpha}
(x,y,t=0)$ basis wave functions is a solution to the time-dependent
Schr\"odinger equation, so each has a uniquely specified time evolution.
Thus,
given (\ref{K93}), we see that the probability integral $\int
dx\,dy\,\hat{\psi}^{\rm
L}(x, y,t)\hat{\psi}^{\rm R}(x,y,t)$ is preserved in time, with all the
terms that
involve powers of $t$ dropping out of $\hat{N}(\hat{\psi},t)$. Therefore, the
equal-frequency theory is unitary.

To summarize the nature of completeness in the Jordan-block case, where there
are fewer energy eigenstates than the dimensionality of the Hamiltonian, the
missing eigenstates are replaced by an equal number of nonstationary
solutions
to the Schr\"odinger equation. Together the stationary and nonstationary
states
form a complete basis that may be used to construct an initial wave packet.
Because these states are complete, the normalization of the wave packet is
preserved in time.

Terms involving powers of $t$ contribute to the Green's functions of the
equal-frequency theory even though they play no role in probability integrals
such as $\hat{N}(\hat{\psi},t)$. We construct the equal-frequency Green's
function $\hat{G}(x,y,x',y',t)$ as the limit of the unequal-frequency Green's
function $G(x,y,x',y',t)$ given in (\ref{K81}). In the unequal-frequency
theory
the representative one-particle contribution to (\ref{K81}) is
\begin{eqnarray}
&&G(x,y,x',y',t)_{(1)}\nonumber\\
&&=\psi^{\rm R}_1(x,y,t=0)\frac{\cC_1}{|N_1|}e^{-iE_1t}\psi_1^{\rm
L}(x',y',t=0)
+\psi^{\rm R}_2(x,y,t=0)\frac{\cC_2}{|N_2|}e^{-iE_2t}\psi_2^{\rm
L}(x',y',t=0)
\nonumber\\
&&=\psi_0^{\rm R}(x,y,t=0)\psi_0^{\rm L}(x',y',t=0)e^{-i(E_0+\omega)t}\frac{4
\gamma^2\omega^2(\omega^2-\epsilon^2)^{1/2}}{\pi\epsilon}\left[e^{-i\epsilon
t}
(\omega+\epsilon)\right.\nonumber\\
&&\times\left.(x'-\omega y'+\epsilon y')(x+\omega y-\epsilon
y)-e^{i\epsilon t}
(\omega-\epsilon)(x'-\omega y'-\epsilon y')(x+\omega y+\epsilon y)\right].
\label{K97}
\end{eqnarray}
Despite the presence of terms that behave as $1/\epsilon$, the $\epsilon\to0$
limit of (\ref{K97}) exists:
\begin{eqnarray}
G(x,y,x',y',t)_{(1)}&\to&\hat{\psi}^{\rm R}_0(x,y,t=0)\hat{\psi}_0^{\rm
L}(x',
y',t=0)e^{-2i\omega t}\frac{8\gamma^2\omega^3}{\pi}\nonumber\\
&&\times\left[(1-i\omega t)(x'-\omega y')(x+\omega y)+\omega y'(x+\omega y)
-\omega (x'-\omega y')y\right],
\label{K98}
\end{eqnarray}
and using (\ref{K85}) and (\ref{K88}), we rewrite (\ref{K98}) as
\begin{eqnarray}
\hat{G}(x,y,x',y',t)_{(1)}&=&\frac{8\gamma^2\omega^3}{\pi}\left[\hat{\psi}^{\rm
R}_{1}(x,y,t)\hat{\psi}^{\rm L}_{1}( x',y',t=0)-\omega\hat{\psi}^{\rm
R}_{1a}(x,
y,t)\hat{\psi}^{\rm L}_{1}(x',y',t=0)\right.\nonumber\\
&&\left.-\omega\hat{\psi}^{\rm R}_{1}(x,y,t)\hat{\psi}^{\rm
L}_{1a}(x',y',t=0)
\right].
\label{K99}
\end{eqnarray}

Equation (\ref{K99}) reveals the role played by the nonstationary states
in the
propagator. Because $\hat{G}(x,y,x',y',t)$ describes the propagation of a
wave
packet that is localized at $(x,y)$ at an initial time $t$, both
stationary and
nonstationary wave functions are contained in the wave packet, and both are
needed to form a complete basis with which to construct localized wave
packets
\cite{R30}.

When the initial and final states are at the same time, the Green's function
gives the normalization of the eigenstates of the position operator
according to
$\langle
x,y,t|x^{\prime},y^{\prime},t\rangle=\delta(x-x^{\prime})\delta(y-y^{
\prime})$. Consequently, the $\epsilon\to0$ limit of the unequal-frequency
completeness relation given in (\ref{K78}) must recover this property of
(\ref{K99}). Explicit evaluation of (\ref{K78}) yields
\begin{eqnarray}
&&\frac{8\gamma^2\omega^3}{\pi}\left[\hat{\psi}^{\rm
R}_{1}(x,y,t)\hat{\psi}^{
\rm L}_{1}( x',y',t)-\omega\hat{\psi}^{\rm R}_{1a}(x,y,t)\hat{\psi}^{\rm
L}_{1}
(x',y',t)-\omega\hat{\psi}^{\rm R}_{1}(x,y,t)\hat{\psi}^{\rm L}_{1a}(x',y',t)
\right]\nonumber\\
&&+\frac{16\gamma^3\omega^4}{\pi}\left[\hat{\psi}^{\rm
R}_{2}(x,y,t)\hat{\psi}^{
\rm L}_{2}(x',y',t)-\omega\hat{\psi}^{\rm R}_{2a}(x,y,t)\hat{\psi}^{\rm
L}_{2}(
x',y',t)-\omega\hat{\psi}^{\rm R}_{2}(x,y,t)\hat{\psi}^{\rm L}_{2a}(x',y',t)
\right]\nonumber\\
&&+\frac{8\gamma^3\omega^6}{\pi}\left[\hat{\psi}^{\rm
R}_{2a}(x,y,t)\hat{\psi}^{
\rm L}_{2a}(x',y',t)-\hat{\psi}^{\rm R}_{2b}(x,y,t)\hat{\psi}^{\rm
L}_{2}(x',y',
t)-\hat{\psi}^{\rm R}_{2}(x,y,t)\hat{\psi}^{\rm
L}_{2b}(x',y',t)\right]+\ldots
\nonumber\\
&&=\delta(x-x^{\prime})\delta(y-y^{\prime})
\label{K100}
\end{eqnarray}
at all times, just as required. Using (\ref{K95}) and the orthogonality of
the
different multiparticle wave functions, one can verify (\ref{K100}) by
projecting with $\int dx\,dy\,\hat{\psi}^{\rm L}_1(x,y,t)$, and so on.
Equation
(\ref{K100}) thus generalizes the standard completeness relation to the
nonstationary case.

We have shown in this section how to develop a consistent quantum-mechanical
theory given that the energy eigenstates do not form a complete basis. We did
not make use of the fact that energy is quantized. Rather we introduced a
specific canonical form for the commutators of the position and momentum
operators, and we did this without reference to the structure of the
Hamiltonian. Demanding that position and momentum operators such as $x$, $p$,
$y$, and $q$ be quantum operators required that we specify a Hilbert space on
which they operate. In the basis in which the position operators are
diagonal,
we introduced a complete set of basis vectors $|x,y\rangle$, and in so
doing we
specified the Hilbert space once and for all.

An alternative but equivalent prescription is to represent the position and
momentum operators as infinite-dimensional matrices acting on a Fock space,
where all the states are created from a no-particle state. The raising
operators
then generate the complete set of basis vectors. This procedure works
whether or
not the Hamiltonian commutes with the Fock-space number operator. This
construction does not involve the Hamiltonian, and it does not depend on how
many energy eigenstates the Hamiltonian possesses or on which states in
the Fock
space are its eigenvectors.

Normally, in quantum mechanics the Hamiltonian is a Hermitian operator whose
energy eigenstates form a complete basis. This basis is in one-to-one
correspondence with both the coordinate-space basis and the Fock-space basis.
Hermitian Hamiltonians can be diagonalized, so for a Hermitian Hamiltonian
it is
advantageous to use its eigenvectors rather than any other set of vectors
as the
basis vectors. In this paper we have shown that one need not have a complete
basis of energy eigenstates to characterize a quantum-mechanical Hilbert
space
when the Hamiltonian is in Jordan-block form. If the Hamiltonian cannot be
diagonalized, the coordinate-space and Fock-space bases are central and the
nonstationary solutions to the Schr\"odinger equation play a role that
they do
not play in the case of a Hermitian Hamiltonian. The lack of
diagonalizability
of Jordan-block Hamiltonians is not an impediment to the construction of a
fully
consistent and unitary Jordan-block quantum theory, and for such cases
$\cP\cT$
quantum mechanics represents a distinct realization of quantum mechanics that
exists in and of itself.

\section{Conclusions and Comments}
\label{s7}

The fourth-order-derivative Pais-Uhlenbeck model is rich and instructive
and we
have used it to examine many issues of contemporary concern. In our previous
paper \cite{R5} we showed that the unequal-frequency version of the model
has a
consistent quantum realization in which the spectrum is real and bounded
below,
the Hilbert space of states is ghost free, and time evolution is unitary. In
this paper we have examined the case of the equal-frequency Pais-Uhlenbeck
model. We have constructed this model by performing the equal-frequency
limit of
the unequal-frequency model. This limit is singular because the Hamiltonian
develops a Jordan-block structure and many of the eigenstates of the
Hamiltonian
disappear. Nevertheless, we have shown that the limiting theory remains a
consistent and unitary quantum theory.

Our solution to the equal-frequency Pais-Uhlenbeck model stems from our
work on
the unequal-frequency model. To find a physically acceptable realization
of the
unequal-frequency model we established in Ref.~\cite{R5} that the
Pais-Uhlenbeck
model Hamiltonian $H$ is actually not Dirac Hermitian, but is instead
$\cP\cT$
symmetric. This symmetry allowed us to construct a similarity transformation
${\tilde H}=e^{-\cQ/2}He^{\cQ/2}$, which produces the Dirac Hermitian
Hamiltonian $\tilde H$ having the same eigenvalues as $H$. We calculated the
operator $\cQ$ exactly and in closed form. The $\cQ$ operator reveals the
singular nature of the equal-frequency limit because it ceases to exist in
this
limit. Thus, there is no equivalent Dirac-Hermitian Hamiltonian for the
equal-frequency theory. Nevertheless, we have shown in this paper that all of
the eigenfunctions that disappear in the equal-frequency limit are
replaced by
time-dependent solutions to the Schr\"odinger equation. As a result,
completeness is maintained and the model continues to exhibit unitary time
evolution. Thus, we have shown that the equal-frequency Pais-Uhlenbeck
model is
a unitary $\cP\cT$ quantum theory that has no equivalent Hermitian
counterpart, and so $\cP\cT$ quantum mechanics should be regarded as being
on an
equal footing with standard Hermitian quantum mechanics, and as being
completely
independent of it at the special critical points where the operator $\cQ$ is
singular.

Having shown that the Pais-Uhlenbeck Hamiltonian defines a physically
acceptable
quantum-mechanical theory, we believe that the techniques and results that we
have described here will be of value in quantum field theory. The original
motivation of Pais and Uhlenbeck was to see if one could avoid the
renormalization infinities in theories such as quantum electrodynamics by
having
Feynman propagators that behave as $1/k^4$ rather than as $1/k^2$. The
discouraging result that they found was that one could do so but at the
price of
having an energy spectrum without a lower bound. Subsequently, following the
development of indefinite-metric theories, it was realized that one could
evade
this problem and have a spectrum that is bounded below, but one would have to
pay a different and equally unpalatable price, namely, allowing states of
negative Dirac norm and violating the physical requirement of unitarity. 
In our
work on the Pais-Uhlenbeck model we have been able to overcome both the
spectral
nonpositivity and the norm nonpositivity (ghost) problems. Our optimism that
ghost problems in quantum field theory can be solved is strengthened by
the fact
that $\cP\cT$ techniques have previously been used \cite{R11} to show that
the
quantum-field-theoretic Lee model is ghost free.

There are many possible directions for future research. Perhaps, the most
intriguing future application of the ideas developed in this paper is to
attempt
to construct a consistent fourth-order derivative quantum theory of
gravity in
four spacetime dimensions. (Success in this endeavor might eliminate the need
for ten-dimensional string theory.) Quantum field theory is beyond the
scope of
this paper, but we note that the Pais-Uhlenbeck model serves as a
quantum-mechanical prototype for higher-derivative theories, such as
conformal
gravity \cite{R31}, which seek to construct a consistent quantum theory of
gravity. Roughly speaking, the issues involved are illustrated by the
fourth-order scalar field theory whose action is
\medskip
\begin{equation}
I=\frac{1}{2}\int d^4x\left[\partial_{\mu}\partial_{\nu}\phi\partial^{\mu}
\partial^{\nu}\phi-(M_1^2+M_2^2)\partial_{\mu}\phi\partial^{\mu}\phi+M_1^2M_2^2
\phi^2\right].
\label{K101}
\end{equation}
Here, the scalar field $\phi$ can be thought of as representing a typical
component of the metric fluctuation $h_{\mu\nu}$ around a flat background
$\eta_{\mu\nu}$. The action (\ref{K101}) gives the equation of motion
\begin{equation}
(\partial_t^2-\nabla^2+M_1^2)(\partial_t^2-\nabla^2+M_2^2)\phi(\bar{x},t)=0.
\label{K102}
\end{equation}
With $k^2=k_0^2-\bar{k}^2$, the propagator for this theory is given by
\begin{eqnarray}
&&(\partial_t^2-\nabla^2+M_1^2)(\partial_t^2-\nabla^2+M_2^2)D^{(4)}(x-y)=
\delta^4(x-y),\nonumber\\
&&D^{(4)}(k^2)=\frac{1}{(k^2-M_1^2)(k^2-M_2^2)}=\frac{1}{M_1^2-M_2^2}\left(
\frac{1}{k^2-M_1^2}-\frac{1}{k^2-M_2^2}\right).
\label{K103}
\end{eqnarray}
This propagator  exhibits the good $1/k^4$ convergence at large $k^2$ that is
needed to make the quantum theory renormalizable, but it appears to do so
at the
expense of having ghost states. The identifications
$\omega_1=(\bar{k}^2+M_1^2
)^{1/2}$, $\omega_2=(\bar{k}^2+M_2^2)^{1/2}$ reduce (\ref{K103}) to
(\ref{K29}),
so there is hope that there might be a solution to the higher-derivative
gravity
ghost problem that parallels the solution to the Pais-Uhlenbeck-model ghost
problem presented here and in Ref.~\cite{R5}.

To see how things could possibly develop, we note that for the generic
higher-derivative scalar action $I=\int d^4x\,{\cal
L}(\phi,\partial_{\mu}\phi,
\partial_{\mu}\partial_{\nu}\phi)$ the translation invariance of the theory
leads to an energy-momentum tensor of the form
\begin{equation}
T_{\mu\nu}=\left[\frac{\partial{\cal L}}{\partial
\phi_{,\mu}}-\partial_{\lambda
}\left(\frac{\partial {\cal L}}{\partial\phi_{,\mu,\lambda}}\right)\right]
\phi_{,\nu}+\frac{\partial {\cal L}}{\partial \phi_{,\mu,\lambda}}\phi_{,\nu,
\lambda}-\eta_{\mu\nu}{\cal L}.
\label{K104}
\end{equation}
This energy-momentum tensor is conserved in field configurations that obey
the
fourth-order-derivative Euler-Lagrange equation
\begin{equation}
\frac{\partial{\cal L}}{\partial\phi}-\partial_{\mu}\left(\frac{\partial{\cal
L}}{\partial \phi_{,\mu}}\right)+\partial_{\mu}\partial_{\nu}\left(\frac{
\partial {\cal L}}{\partial\phi_{,\mu,\nu}}\right)=0.
\label{K105}
\end{equation}

In a phase space formulation of the theory, $T_{\mu\nu}$ is replaced by
\begin{equation}
T_{\mu\nu}=\pi_{\mu}\phi_{,\nu}+\pi_{\mu\lambda}\phi_{,\nu,\lambda}-\eta_{\mu\nu
}{\cal L}.
\label{K106}
\end{equation}
Thus, for the particular action in (\ref{K101}) $T_{\mu\nu}$ has the form
\begin{equation}
T_{\mu\nu}=\pi_{\mu}\phi_{,\nu}+\pi_{\mu\lambda}\pi^{\lambda}_{\phantom{\lambda}
\nu}-\frac{1}{2}\eta_{\mu\nu}\left[\pi_{\lambda\kappa}\pi^{\lambda\kappa}-(M_1^2
+M_2^2)\partial_{\lambda}\phi\partial^{\lambda}\phi+M_1^2M_2^2\phi^2\right],
\label{K107}
\end{equation}
and the equation of motion is (\ref{K102}). For the metric
$\eta_{\mu\nu}=(1,-1,
-1,-1)$, the Hamiltonian of the theory is $H=\int d^3x\,T_{00}$, where
\begin{equation}
T_{00}=\pi_{0}\dot{\phi}+\frac{1}{2}\pi_{00}^2+\frac{1}{2}(M_1^2+M_2^2)\dot{
\phi}^2-\frac{1}{2}M_1^2M_2^2\phi^2-\frac{1}{2}\pi_{ij}\pi^{ij}+\frac{1}{2}(
M_1^2+M_2^2)\phi_{,i}\phi^{,i}.
\label{K108}
\end{equation}

We note that (\ref{K108}) is similar in structure to $H_{\rm PU}$ in
(\ref{K30}), and (\ref{K108}) is its covariant generalization. Recalling the
identifications $\omega_1=(\bar{k}^2+M_1^2)^{1/2}$,
$\omega_2=(\bar{k}^2+M_2^2
)^{1/2}$, the Hamiltonian (\ref{K45}) and commutation relations (\ref{K46})
covariantly generalize to
\begin{eqnarray}
H&=&\int d^3k\,\bigg{[}2(M_1^2-M_2^2)(\bar{k}^2+M_1^2)\hat{a}_{1,\bar{k
}}a_{1,\bar{k}}+2(M_1^2-M_2^2)(\bar{k}^2+M_2^2)\hat{a}_{2,\bar{k}}a_{2,
\bar{k}}\nonumber\\
&&+\frac{1}{2}(\bar{k}^2+M_1^2)^{1/2}+\frac{1}{2}(\bar{k}^2+M_2^2)^{1/2}\bigg{]}
,\nonumber\\
&&[a_{1,\bar{k}},\hat{a}_{1,\bar{k}^{\prime}}]=[2(M_1^2-M_2^2)(\bar{k}^2+
M_1^2)^{1/2}]^{-1}\delta^3(\bar{k}-\bar{k}^{\prime}),\nonumber\\
&&[a_{2,\bar{k}},\hat{a}_{2,\bar{k}^{\prime}}]=[2(M_1^2-M_2^2)(\bar{k}^2+
M_2^2)^{1/2}]^{-1}\delta^3(\bar{k}-\bar{k}^{\prime}),\nonumber\\
&&[a_{1,\bar{k}},a_{2,\bar{k}^{\prime}}]=0,\quad[a_{1,\bar{k}},\hat{a}_{2,
\bar{k}^{\prime}}]=0,\quad[\hat{a}_{1,\bar{k}},a_{2,\bar{k}^{\prime}}]=0,\quad
[\hat{a}_{1,\bar{k}},\hat{a}_{2,\bar{k}^{\prime}}]=0.
\label{K109}
\end{eqnarray}
In (\ref{K109}) all relative signs are positive.

Similarly, in the limiting case where $M_1^2=M_2^2=M^2$, the above
equations are
replaced by the covariant generalizations of the Jordan-block Hamiltonian
(\ref{K54}) and the commutation relations (\ref{K56}):
\begin{eqnarray}
H&=&\int d^3k\,\bigg{[}8(\bar{k}^2+M^2)^2[2\hat{b}_{\bar{k}}b_{\bar{k}}
+\hat{a}_{\bar{k}}b_{\bar{k}}+\hat{b}_{\bar{k}}a_{\bar{k}}]+(\bar{k}^2+M^2)^{1/
2}\bigg{]},\nonumber\\
&&[a_{\bar{k}},\hat{b}_{\bar{k}^{\prime}}]=[b_{\bar{k}},\hat{a}_{\bar{k}^{\prime
}}]=[8(\bar{k}^2+M^2)^{3/2}]^{-1}\delta^3(\bar{k}-\bar{k}^{\prime}),
\nonumber\\
&&[a_{\bar{k}},\hat{a}_{\bar{k}^{\prime}}]=0,\quad[b_{\bar{k}},\hat{b}_{\bar{k
}^{\prime}}]=0,\quad[a_{\bar{k}},b_{\bar{k}^{\prime}}]=0,\quad[\hat{a}_{\bar{k}}
,\hat{b}_{\bar{k}^{\prime}}]=0.
\label{K110}
\end{eqnarray}
Now, there are zero-norm states and this continues to be the case even if
we set
$M^2=0$. Thus, we find zero-norm states in the pure fourth-order conformal
gravity case where the action can schematically be represented by
$I=\frac{1}{2}
\int d^4x\,\partial_{\mu}\partial_{\nu}\phi\partial^{\mu}\partial^{\nu}\phi$
alone \cite{R32}.

C.~M.~B. is supported by a grant from the U.S. Department of Energy.

\appendix
\setcounter{equation}{0}
\def\theequation{A\arabic{equation}}

\section{Lehmann Spectral Representation for Higher-Derivative Theories}

To derive the Lehmann spectral representation in a higher-derivative field
theory, we follow the same procedure as that for a second-order derivative
field
theory. The derivation makes use of three assumptions: the Poincar\'e
transformation properties of the interacting fields, a completeness
relation for
the exact energy eigenstates of the interacting theory, and the existence
of a
stable ground state $|\Omega\rangle$ whose four-momentum we can take to be
zero.

For the case of a self-interacting Hermitian scalar field $\phi(x)$, one
introduces the translation generator $P_{\mu}=\int d^3xT_{0\mu}$, with the
scalar field then transforming according to
\begin{equation}
[\phi(x),P_{\mu}]=i\partial_{\mu}\phi(x),\qquad \phi(x)=e^{iP\cdot
x}\phi(0)e^{-
iP\cdot x}.
\label{A1}
\end{equation}
Given (\ref{A1}), the matrix element of $\phi(x)$ between the vacuum and the
one-particle state of four-momentum $k^n_{\mu}$, positive $k^n_0$, and
squared
mass $m_n^2=k^n\cdot k^n $ is:
\begin{equation}
\langle \Omega |\phi(x)|k^n_{\mu}\rangle =\langle \Omega
|\phi(0)|k^n_{\mu}\rangle e^{-ik^n\cdot x},\qquad
\langle k^n_{\mu} |\phi(x)|\Omega \rangle =\langle k^n_{\mu}
|\phi(0)|\Omega\rangle e^{ik^n\cdot x}.
\label{A2}
\end{equation}

Let us provisionally take the completeness relation for the four-momentum
eigenstates to be of the conventional Dirac form
\begin{equation}
\sum_n|n\rangle\langle n|={\bf 1}.
\label{A3}
\end{equation}
We can then write the two-point function of the field $\phi(x)$ in the form
\begin{equation}
\langle\Omega|\phi(x)\phi(y)|\Omega\rangle=\sum_n|\langle\Omega|\phi(0)|k^n_{
\mu}\rangle|^2e^{-ik^n\cdot(x-y)}.
\label{A4}
\end{equation}
Introducing the spectral function $\rho(q^2)$ defined by
\begin{equation}
\rho(q^2)=(2\pi)^3\sum_n\delta^4(k^n_{\mu}-q_{\mu})|\langle\Omega
|\phi(0)|k^n_{\mu}\rangle|^2\theta(q_0),
\label{A5}
\end{equation}
we write the two-point function as
\begin{equation}
\langle\Omega|\phi(x)\phi(y)|\Omega\rangle=\int_0^{\infty}dm^2\,\rho(m^2)\int
\frac{d^4q}{(2\pi)^3}\theta(q_0)\delta(q^2-m^2)e^{-iq\cdot(x-y)}.
\label{A6}
\end{equation}

Repeating the same analysis for the
$\langle\Omega|\phi(y)\phi(x)|\Omega\rangle$
two-point function, we obtain the usual Lehmann representation
\begin{eqnarray}
\Delta_{\rm F}^{\rm int}(x-y)=i\langle \Omega
|T[\phi(x)\phi(y)]|\Omega\rangle
&=&i\langle\Omega|\left[\theta(x_0-y_0)\phi(x)\phi(y)+\theta(y_0-x_0)\phi(y)
\phi(x)\right]|\Omega\rangle\nonumber\\
&=& \int_0^{\infty}dm^2\,\rho(m^2)\Delta_{(\rm F,2)}^{\rm free}(x-y;m^2),
\label{A7}
\end{eqnarray}
for the time-ordered product of the interacting fields \cite{R19}. Here,
$\Delta
_{(\rm F,2)}^{\rm
free}(x-y;m^2)=[i\langle\Omega|T[\phi(x)\phi(y)]|\Omega\rangle
]_{\rm free}$ is the Feynman propagator for a free scalar field with mass
$m$,
namely the  propagator that obeys the free second-order-theory relations
\begin{equation}
\left[\partial_t^2-\nabla^2+m^2\right]\Delta_{(\rm F,2)}^{\rm free}(x-y;m^2)=
\delta^4(x-y),\qquad\Delta^{\rm free}_{(\rm
F,2)}(k^2;m^2)=-\frac{1}{k^2-m^2}.
\label{A8}
\end{equation}
(Comparing (\ref{A8}) with the fourth-order propagator in (\ref{K103}), we
note a relative minus sign in Fourier space.)

Equation (\ref{A7}) expresses the exact Feynman propagator of the interacting
theory as a spectral integral over free second-order Feynman propagators
with a
continuum of mass values. No assumption has been made regarding the order
of the
equation of motion obeyed by the interacting field $\phi(x)$.
Nevertheless, the
free propagator $\Delta_{(\rm F,2)}^{\rm free}(x-y;m^2)$ in (\ref{A7}) still
satisfies a second-order differential equation. Independent of the
structure or
order of the interacting field equation, the mass-shell condition associated
with the eigenstates of the exact four-momentum operator $P_{\mu}$ is still a
second-order condition because of Poincar\'e invariance.

The derivation of (\ref{A7}) is generic, and we can apply it to a field
theory
having a fourth-order field equation. To do so we need to show that in the
absence of interactions the quantity $i\langle\Omega|T[\phi(x)\phi(y)]|\Omega
\rangle$ can indeed be identified with the free fourth-order propagator
\begin{eqnarray}
D^{(4)}(k^2)=\frac{1}{M_1^2-M_2^2}\left(
\frac{1}{k^2-M_1^2}-\frac{1}{k^2-M_2^2}\right)
\label{A9}
\end{eqnarray}
introduced above. In the free fourth-order case associated with the
equation of
motion
\begin{equation}
(\partial_t^2-\nabla^2+M_1^2)(\partial_t^2-\nabla^2+M_2^2)\phi(\bar{x},t)=0,
\label{A10}
\end{equation}
we must evaluate the relevant time derivatives of the two-field T-product.

For the first time derivative we obtain
\begin{eqnarray}
\frac{\partial}{\partial x_0}\left[\langle \Omega
|T[\phi(x)\phi(y)]|\Omega
\rangle\right]&=&\langle\Omega|\left[\theta(x_0-y_0)
\dot{\phi}(x)\phi(y)+\theta(y_0-x_0)\phi(y)\dot{\phi}(x)\right]|\Omega\rangle
\nonumber\\
&+&\langle\Omega |\delta(x_0-y_0)[\phi(x),\phi(y)]|\Omega \rangle,
\label{A11}
\end{eqnarray}
and, with $\delta(x_0-y_0)[\phi(x),\phi(y)]=0$, we obtain
\begin{eqnarray}
\frac{\partial}{\partial
x_0}\left[\langle\Omega|T[\phi(x)\phi(y)]|\Omega\rangle
\right]&=&\langle\Omega|\left[\theta(x_0-y_0)\dot{\phi}(x)\phi(y)+\theta(y_0-x_0
)\phi(y)\dot{\phi}(x)\right]|\Omega\rangle.
\label{A12}
\end{eqnarray}
Analogously, for the second time derivative we obtain
\begin{eqnarray}
\frac{\partial^2}{\partial x_0^2}\left[\langle\Omega
|T[\phi(x)\phi(y)]|\Omega\rangle\right]&=&\langle\Omega|\left[\theta(x_0-y_0)
\ddot{\phi}(x)\phi(y)+\theta(y_0-x_0)\phi(y)\ddot{\phi}(x)\right]|\Omega\rangle
\nonumber\\
&+&\langle\Omega|\delta(x_0-y_0)[\dot{\phi}(x),\phi(y)]|\Omega \rangle.
\label{A13}
\end{eqnarray}
However, unlike the second-order case, $\phi(x)$ and $\dot{\phi}(x)$ are not
canonical conjugates. Rather, they are the covariant analogs of the $y$ and
$\dot{y}=i[H,y]=-ix$ variables of the fourth-order Pais-Uhlenbeck oscillator
theory. With the $y$ and $x$ operators being commuting variables, the
fourth-order equal-time commutator
$\delta(x_0-y_0)[\dot{\phi}(x),\phi(y)]$ must
thus vanish. Consequently, (\ref{A13}) reduces to
\begin{equation}
\frac{\partial^2}{\partial x_0^2}\left[\langle\Omega|T[\phi(x)\phi(y)]|\Omega
\rangle\right]=\langle\Omega|\left[\theta(x_0-y_0)\ddot{\phi}(x)\phi(y)+\theta
(y_0-x_0)\phi(y)\ddot{\phi}(x)\right]|\Omega \rangle.
\label{A14}
\end{equation}

For the third time derivative we obtain
\begin{eqnarray}
\frac{\partial^3}{\partial x_0^3}\left[\langle\Omega|T[\phi(x)\phi(y)]|\Omega
\rangle\right]&=&\langle\Omega|\left[\theta(x_0-y_0)\frac{\partial^3\phi(x)}{
\partial
x_0^3}\phi(y)+\theta(y_0-x_0)\phi(y)\frac{\partial^3\phi(x)}{\partial
x_0^3}\right]|\Omega\rangle\nonumber\\
&&\quad+\langle\Omega|\delta(x_0-y_0)[\ddot{\phi}(x),\phi(y)]|\Omega\rangle.
\label{A15}
\end{eqnarray}
Because $\phi(x)$ and $\ddot{\phi}(x)$ are the covariant analogs of the
$y$ and
$-i\dot{x}=[H,x]=-ip/\gamma$ variables, the commutator
$\delta(x_0-y_0)[\ddot{
\phi}(x),\phi(y)]$ also vanishes. Thus, (\ref{A15}) reduces to
\begin{eqnarray}
\frac{\partial^3}{\partial x_0^3}\left[\langle\Omega|T[\phi(x)\phi(y)]|\Omega
\rangle\right]&=&\langle\Omega|\left[\theta(x_0-y_0)\frac{\partial^3\phi(x)}{
\partial
x_0^3}\phi(y)+\theta(y_0-x_0)\phi(y)\frac{\partial^3\phi(x)}{\partial
x_0^3}\right]|\Omega\rangle.\nonumber\\
\label{A16}
\end{eqnarray}

Finally, for the fourth time derivative we obtain
\begin{eqnarray}
\frac{\partial^4}{\partial x_0^4}\left[\langle\Omega|T[\phi(x)\phi(y)]|\Omega
\rangle\right]&=&\langle\Omega|\left[\theta(x_0-y_0)\frac{\partial^4\phi(x)}{
\partial
x_0^4}\phi(y)+\theta(y_0-x_0)\phi(y)\frac{\partial^4\phi(x)}{\partial
x_0^4}\right]|\Omega\rangle\nonumber\\
&&\quad+\langle\Omega|\delta(x_0-y_0)\left[\frac{\partial^3\phi(x)}{\partial
x_0^3},\phi(y)\right]|\Omega\rangle.
\label{A17}
\end{eqnarray}
However, $\partial^3\phi(x)/\partial x_0^3$ is the analog of
$-i\dot{p}/\gamma=
[H,p]/\gamma=i(\omega_1^2+\omega_2^2)x+q/\gamma$, and the commutator of
$y$ and
$q$ is $[y,q]=i$. Thus, the equal-time commutator $\delta(x_0-y_0)[\partial^3
\phi(x)/\partial x_0^3,\phi(y)]$ is equal to $-i\delta^4(x-y)$. Hence, the
$i$ times the T-product obeys (\ref{K103}), and
$i\langle\Omega|T[\phi(x)\phi(y)]|\Omega\rangle$ is the fourth-order-theory
Green's function we need.

Note that in the fourth-order case in Fourier space the left side of
(\ref{A7})
behaves as $1/k^4$ at large $k^2$. However, the Fourier transform of
$\Delta_{(
\rm F,2)}^{\rm free}(x-y;m^2)$ behaves as $1/k^2$. The spectral function
$\rho(
m^2)$ in (\ref{A5}) is positive definite. Since (\ref{A7}) is a mathematical
identity, we thus have a contradiction. The difference between the large
$k^2$
behaviors of $\Delta_{(\rm F,4)}^{\rm free}(x-y)$ and of $\Delta_{(\rm
F,2)}^{
\rm free}(x-y)$ implies that the spectral function $\rho(m^2)$ is not
positive
definite. Hence, the standard Dirac completeness relation in (\ref{A3})
for the
energy eigenstates cannot be valid in the fourth-order case. This analysis
immediately generalizes to all higher-order derivative theories for which the
propagator is even more convergent at large $k^2$. We conclude that before
constructing the Hilbert space appropriate to a higher-derivative theory, we
know from the outset that the needed inner product cannot be the standard
Dirac
one.

A simple modification of the Dirac norm that gives a nonpositive definite
$\rho(m^2)$ would be to replace (\ref{A3}) by
\begin{equation}
\sum_n\eta_n|n\rangle\langle n|={\bf 1},
\label{A18}
\end{equation}
where $\eta_n$ is $\pm1$, and to replace (\ref{A5}) by
\begin{equation}
\rho(q^2)=(2\pi)^3\sum_n\delta^4(k^n_{\mu}-q_{\mu})|\langle\Omega
|\phi(0)|k^n_{\mu}\rangle|^2\eta_n\theta(q_0).
\label{A19}
\end{equation}
However, this choice would violate unitarity.

When the Hamiltonian is not Hermitian, positivity of the spectral weight
function is no longer mandatory \cite{R33}. However, we need not forego
unitarity because we no longer use the Dirac inner product. Instead, we
use the $\cP\cT$ inner product
\begin{equation}
\langle n|e^{-\cQ}|m\rangle=\delta_{m,n},\qquad
\sum_n|n\rangle\langle n|e^{-\cQ}={\bf 1}
\label{A20}
\end{equation}
introduced earlier. With such an inner product the spectral function of
(\ref{A5}) is
replaced by
\begin{equation}
\rho(q^2)=(2\pi)^3\sum_n\delta^4(k^n_{\mu}-q_{\mu})\langle\Omega|e^{-\cQ}\phi(0)
|k^n_{\mu}\rangle\langle k^n_{\mu}|e^{-\cQ}\phi(0)|\Omega\rangle\theta(q_0).
\label{A21}
\end{equation}

Unlike the spectral function associated with a Dirac norm, this spectral
function need not be positive definite. To evaluate it, in analogy to
(\ref{K44}) we set
\begin{equation}
\phi(x)=\int \frac{d^3k}{(2\pi)^{3/2}}\,\left [-ia_{1,\bar{k}}e^{-ik^1\cdot
x}+a_{2,\bar{k}}e^{-ik^2\cdot x}-i\hat{a}_{1,\bar{k}}e^{ik^1\cdot
x}+\hat{a}_{2,\bar{k}}e^{ik^2\cdot x}\right].
\label{A22}
\end{equation}
We define right and left vacua according to
\begin{equation}
a_{1,\bar{k}}|0_R\rangle =0,\qquad a_{2,\bar{k}}|0_R\rangle =0,\qquad
\langle 0_L|\hat{a}_{1,\bar{k}}=0,\qquad \langle
0_L|\hat{a}_{2,\bar{k}}=0,
\label{A23}
\end{equation}
and in terms of the notation in (\ref{A23}) identify
\begin{eqnarray}
|\Omega \rangle=|0_R\rangle,\qquad
&&|k_{\mu}^{i}\rangle=[2(M_1^2-M_2^2)(\bar{k}^2+
M_i^2)^{1/2}]^{1/2}\hat{a}_{i,\bar{k}}|0_R\rangle,
\nonumber \\
\langle \Omega|e^{-\cQ}=\langle 0_L|,\qquad &&\langle
k_{\mu}^{i}|=[2(M_1^2-M_2^2)(\bar{k}^2+
M_i^2)^{1/2}]^{1/2}\langle 0_L|a_{i,\bar{k}},
\label{A24}
\end{eqnarray}
as normalized according to
\begin{equation}
\langle 0_L|0_R\rangle =1,\qquad \langle
k_{\mu}^{1}|k_{\mu}^{1}\rangle=1,\qquad
\langle k_{\mu}^{2}|k_{\mu}^{2}\rangle=1.
\label{A25}
\end{equation}
Through use of the commutation relations (\ref{K109}) we obtain the
on-shell contributions
\begin{eqnarray}
&&\langle \Omega|e^{-\cQ}\phi(0)|k_{\mu}^1\rangle \langle
k_{\mu}^1|e^{-\cQ}\phi(0)|\Omega \rangle
=-\frac{1}{(2\pi)^32(\bar{k}^2+M_1^2)^{1/2}(M_1^2-M_2^2)},
\nonumber \\
&&\langle \Omega|e^{-\cQ}\phi(0)|k_{\mu}^2\rangle \langle
k_{\mu}^2|e^{-\cQ}\phi(0)|\Omega \rangle
=\phantom{-}\frac{1}{(2\pi)^32(\bar{k}^2+M_2^2)^{1/2}(M_1^2-M_2^2)},
\label{A26}
\end{eqnarray}
to give us precisely the relative minus sign we require. Recalling the
minus sign in (\ref{A8}), insertion of
(\ref{A26}) in the Lehmann representation then yields the propagator
(\ref{A9}). Hence, in fourth-order theories compatibility of the Lehmann
representation with unitarity is readily achievable \cite{R34}.

{}

\begin{thebibliography}{}
\bibitem{R18} Jordan showed that via a sequence of similarity transformations
any $N$-dimensional square matrix can be brought to a form where it is either
diagonal or in triangular Jordan-block form in which all of the elements
on one
side of the diagonal are zero. Since the nonzero elements on the other
side of
the diagonal do not contribute to the secular equation, the $N$ elements
on the
diagonal of a Jordan-block form matrix are then the $N$ eigenvalues of the
matrix. When all of the eigenvalues are real, a Jordan-block matrix is an
example of a non-Hermitian matrix having eigenvalues that are all real.
(Hermiticity is only a sufficient condition for obtaining real
eigenvalues, not
a necessary one. The eigenvalues of a Hermitian matrix are real, but there is
nothing that requires the eigenvalues of a non-Hermitian matrix to be
complex.
Non-Hermitian matrices may have real eigenvalues.) Because Jordan-block
matrices
cannot be brought to a diagonal form by any further similarity transform,
they
possess fewer than $N$ eigenvectors and the eigenvectors of a Jordan block
matrix do not form a complete basis.
\bibitem{R24} A physically equivalent alternative to making the similarity
transformation in (\ref{K37}) is to keep the original operator basis of
(\ref{K30}) and to take the quantum-mechanical operators $\hat{z}$ and
$\hat{p
}_z$ to be anti-Hermitian on the real-$z$ axis. The $[\hat{z},\hat{p}_z]=i$
commutator would then be realized via $\hat{z}=-iz,~\hat{p}_z=\partial_z$
with
the real coordinate $z$. In this case the ground-state eigenfunction would be
given by $\psi_0(z,x)={\rm exp}\left[-(\gamma/2)(\omega_1+\omega_2)\omega_1
\omega_2z^2+\gamma\omega_1\omega_2zx-(\gamma/2)(\omega_1+\omega_2)x^2\right]$
and it would be well behaved on the real-$z$ and real-$x$ axes. 
Alternatively,
for the unbounded-below energy spectrum associated with the unconventional
propagator $G^{\rm UNC}(t)$ of (\ref{K33}) the state of energy
$E_0^{\prime}=(
\omega_1-\omega_2)/2$ solves the Schr\"odinger equation (\ref{K35}) with
eigenfunction $\psi'_0(z,x)={\rm exp}\left[-(\gamma/2)(\omega_1-\omega_2)
\omega_1\omega_2z^2-i\gamma\omega_1\omega_2zx-(\gamma/2)(\omega_1-\omega_2)x^2
\right]$; this eigenfunction is well behaved on the real-$z$ and real-$x$
axes.
The unbounded-below energy sector can thus be associated with the
quantum-mechanical operators $\hat{z}$ and $\hat{p}_z$, which are
Hermitian on
the real-$z$ axis, while the bounded-below energy sector can be associated
with
the quantum-mechanical operators $\hat{z}$ and $\hat{p}_z$, which are
anti-Hermitian on the real-$z$ axis. The two candidate realizations of the
Pais-Uhlenbeck model correspond to different contour prescriptions in the
complex-energy plane, and they are also distinguished by the differing
domains
of Hermiticity that they assign to the position and momentum operators.
To underscore this point, we recall that for the generic one-dimensional
harmonic oscillator $\hat{H}=\hat{v}^2+\hat{u}^2$, with
$[\hat{u},\hat{v}]=i$,
the energy spectrum is manifestly bounded below if the operators $\hat{u}
=(a+a^{\dag})/\surd{2}$ and $\hat{v}=i(a^{\dag}-a)/\surd{2}$ are Hermitian on
the real-$u$ axis; the state $|\Omega\rangle$ that $a$ annihilates
has energy $E_0=1$ and the normalizable ground-state eigenfunction
$\psi_0(u)=e^{-u^2/2}$. However, for this oscillator the Schr\"odinger
equation also admits a solution associated with the state
$|\Omega^{\prime}\rangle$ that $a^{\dag}$ annihilates, and it has negative
energy $\tilde{E}_0=-1$ and eigenfunction $\tilde{\psi}_0(u)=e^{u^2/2}$.
This eigenfunction is not normalizable on the real-$u$ axis, but is
normalizable on the imaginary-$u$ axis. Moreover, in this sector the
excited state $a|\Omega^{\prime}\rangle$ is an energy eigenstate with a
Dirac norm $\langle \Omega^{\prime}|a^{\dag}a|\Omega^{\prime}\rangle$,
which is negative. The eigenfunctions in this sector would become
normalizable on the real axis and their norms would become positive if the
operators $\hat{u}$ and $\hat{v}$ were anti-Hermitian on the real axis.
Thus obtaining non-normalizable wave functions or negative Dirac norms
serves as a warning that certain operators are not Hermitian on the real
axis. The choice of domain of Hermiticity of the position and momentum
operators determines which of the two realizations (bounded or unbounded
energy spectrum) the harmonic oscillator takes. While all norms are
positive in both of these domains when they are constructed properly, in
only one (Hermitian on the real $u$-axis) is the energy spectrum bounded
below. An analogous situation occurs for the two-oscillator Pais-Uhlenbeck
model except that this time the operators of one of the two oscillators
would need to be anti-Hermitian on the real axis to obtain an energy
spectrum that is  both bounded-below and free of negative norm states.
\bibitem{R25} The correlator $\langle\Omega|e^{-\cQ}y(t)y(0)|\Omega\rangle$
obeys the fourth-order derivative Pais-Uhlenbeck wave equation
(\ref{K27}), but
after the $e^{-\cQ/2}$ transformation the correlator $\langle\tilde{\Omega}|
\tilde{y}(t)\tilde{y}(0)|\tilde{\Omega}\rangle$ will obey the fourth-order
equation too even though $\langle
\tilde{\Omega}|y(t)y(0)|\tilde{\Omega}\rangle$
obeys a second order wave equation. While $H$ and $\tilde{H}$ have the same
energy eigenspectra, the correlators of interest to the Pais-Uhlenbeck theory
are those associated with $H$ rather than those associated with $\tilde{H}$.
\bibitem{R26} As well as deriving (\ref{K45}) and (\ref{K46}) starting from
(\ref{K42}) and (\ref{K43}), we can also derive these relations from the form
for $\tilde{H}$ given in (\ref{K41}). In terms of $\tilde{A}_{\pm}=(2\gamma
\omega_1)^{-1/2}(p\pm i\gamma \omega_1 x)$ and $\tilde{B}_{\pm}=(2\gamma
\omega_1^2\omega_2)^{-1/2}(q\pm i\gamma \omega_1^2\omega_2 y)$, which obey
$[\tilde{A}_-,\tilde{A}_+]=[\tilde{B}_-,\tilde{B}_+]=1$, $[\tilde{A}_{\pm},
\tilde{B}_{\pm}]=[\tilde{A}_{\pm},\tilde{B}_{\mp}]=0$, $\tilde{H}$
diagonalizes
as $\tilde{H}=\omega_1\tilde{A}_+\tilde{A}_-+\omega_2\tilde{B}_+\tilde{B}_-+(
\omega_1+\omega_2)/2$. On transforming according to $A_{\pm}=e^{\cQ/2}\tilde{
A}_{\pm}e^{-\cQ/2}=[2\gamma\omega_1(\omega_1^2-\omega_2^2)]^{-1/2}(\omega_1p+i
\gamma\omega_1\omega_2^2y\pm i\gamma\omega_1^2x \pm q)$, $B_{\pm}=e^{\cQ/2}
\tilde{B}_{\pm}e^{-\cQ/2}=[2\gamma\omega_2(\omega_1^2-\omega_2^2)]^{-1/2}(q+i
\gamma\omega_2^2x\pm i\gamma\omega_1^2\omega_2y \pm \omega_2p)$ and defining
$A_-=-a_1[2\gamma\omega_1(\omega_1^2-\omega_2^2)]^{1/2}$, $A_+=-\hat{a}_1[2
\gamma\omega_1(\omega_1^2-\omega_2^2)]^{1/2}$,
$B_-=-ia_2[2\gamma\omega_2(\omega
_1^2-\omega_2^2)]^{1/2}$,
$B_+=i\hat{a}_2[2\gamma\omega_2(\omega_1^2-\omega_2^2)
]^{1/2}$, (\ref{K45}) and (\ref{K46}) then follow. We thus confirm the
pattern
of coefficients in (\ref{K44}) and the absence of  negative energies in
(\ref{K45}) or ghost-sign commutators in (\ref{K46}).
\bibitem{R28} Appropriately modified, the discussion given here parallels
that
given in \cite{R22}.
\bibitem{R30} The need for the nonstationary
$\hat{\psi}^{\rm}_{1a}(x,y,t)$ for
localization purposes was also noted in \cite{R22}, where the
deep-Euclidean-time limit of the equal-frequency path integral (the
equal-frequency limit of (\ref{K34})) was found to yield an asymptotic
term of
the form $\tau e^{-2\omega\tau}$.
\bibitem{R34} We are indebted to Dr.~A.~M.~Polyakov for suggesting that we
look at the unitarity problem for fourth-order theories  from the
perspective of the Lehmann representation.
\end{thebibliography}
\end{document}